\documentclass[10pt]{elsarticle}

\usepackage{printlen}
\usepackage[english]{babel}
\usepackage[utf8]{inputenc}
\usepackage{hyperref}
\usepackage{verbatim}

\usepackage{graphicx}
\usepackage[caption=false]{subfig}
\usepackage{color}

\usepackage{amsmath,amsfonts,amssymb,amsthm}
\usepackage{mathtools}
\usepackage{empheq}
\usepackage{bm}
\usepackage[
  space-before-unit,
  range-units=repeat,
  detect-weight=true,
  detect-family=true
]{siunitx}

\usepackage{geometry}
\usepackage{pdflscape}

\usepackage{grffile}

\usepackage{multirow}
\usepackage{longtable}

\usepackage{algorithm}
\usepackage[]{algpseudocode}

\usepackage{placeins} 
\usepackage{url}

\usepackage{diagbox}

\usepackage[outline]{contour}

\hypersetup{pdftex,colorlinks=true,allcolors=black}
\usepackage{hypcap}

\usepackage{algorithm}
\usepackage[]{algpseudocode}

\contourlength{2.0pt}

\newcommand{\brc}[1]{\left( #1 \right)}
\newcommand{\sqrbrc}[1]{\left[ #1 \right]}

\newcommand{\avr}[1]{\left<{#1}\right>}
\newcommand{\txt}[1]{\text{#1}}

\newcommand{\nor}[1]{\left\lVert #1 \right\rVert}

\newcommand{\repl}[2]{\mathrlap{#2}\protect\phantom{#1}}

\renewcommand{\vec}[1]{\boldsymbol #1 }

\makeatletter
\newcommand{\pushright}[1]{\ifmeasuring@#1\else\omit\hfill$\displaystyle#1$\fi\ignorespaces}
\newcommand{\pushleft}[1]{\ifmeasuring@#1\else\omit$\displaystyle#1$\hfill\fi\ignorespaces}
\makeatother

\algblock{ParFor}{EndParFor}
\algnewcommand\algorithmicparfor{\textbf{parfor}}
\algnewcommand\algorithmicpardo{\textbf{do}}
\algnewcommand\algorithmicendparfor{\textbf{end\ parfor}}
\algrenewtext{ParFor}[1]{\algorithmicparfor\ #1\ \algorithmicpardo}
\algrenewtext{EndParFor}{\algorithmicendparfor}

\algtext*{EndIf}%
\algtext*{EndFor}%
\algtext*{EndWhile}%
\algtext*{EndParFor}%

\journal{Journal of Computational Physics}

\graphicspath{{Figures/}}

\begin{document}

\begin{frontmatter}

\title{Tensor methods for the Boltzmann-BGK equation}

\author[stanford]{Arnout M. P. Boelens}
\author[ucsc]{Daniele Venturi}
\author[stanford]{Daniel M. Tartakovsky\corref{correspondingAuthor}}
\cortext[correspondingAuthor]{Corresponding author}
\ead{tartakovsky@stanford.edu}

\address[stanford]{Department of Energy Resources Engineering, Stanford University, Stanford, CA 94305}
\address[ucsc]{Department of Applied Mathematics, UC Santa Cruz, Santa Cruz, CA 95064}

\begin{abstract}
We present a tensor-decomposition method to solve the 
Boltzmann transport equation (BTE) in the Bhatnagar-Gross-Krook 
approximation. The method represents the six-dimensional BTE as a 
set of six one-dimensional problems, which are solved with the 
alternating least-squares algorithm and the discrete Fourier transform 
at $N$ collocation points. We use this method to predict the equilibrium 
distribution (steady-state simulation) and a non-equilibrium distribution returning 
to the equilibrium (transient simulation). Our numerical experiments 
demonstrate $N \log N$ scaling. Unlike many BTE-specific numerical 
techniques, the numerical tensor-decomposition method we propose 
is a general technique that can be applied to other high-dimensional 
systems.
\end{abstract}

\begin{keyword}
high-dimensional PDE \sep tensor method \sep non-equilibrium
\end{keyword}

\end{frontmatter}

\section{Introduction}

Whenever the mean free path of molecules becomes larger than the characteristic
length scale of a system, the continuity assumption breaks down and so does the
validity of the Navier-Stokes equations.  This phenomenon occurs in a number of
settings, including splashing droplets~\citep{boelens2018a}, moving contact
lines~\citep{sprittles2017}, super- and hyper-sonic flows~\citep{boyd1995}, and
flow of electrons in metals~\citep{sondheimer1952} and silicon~\citep{jin2008}. 
The physics in this flow regime is often described by the six-dimensional 
(plus time) Boltzmann transport equation (BTE)~\citep{boltzmann1872}. 

Like many other high-dimensional partial differential equations (PDEs), 
the BTE suffers from the curse of dimensionality: the computational
cost of conventional numerical schemes, such as those based on 
tensor product representations, grows exponentially with an increasing 
number of degrees of freedom. One way to mitigate such computational 
complexity is to use particle-based methods~\citep{dimarco2014}, e.g., 
direct simulation Monte Carlo (DSMC)~\citep{bird1963} or the Nambu–Babovsky 
method~\citep{babovsky1986}. These methods preserve the main 
physical properties of the system, even far from equilibrium, and are 
computationally efficient away from near-fluid regimes. In particular, 
they have low memory requirements and their cost scales linearly 
with the number of particles.
However, their accuracy, efficiency and convergence rate tend 
to be poor for non-stationary flows, or flows close to continuum 
regimes \citep{pareschi2001,peraud2014,rogier1994}. 
This is due to the non-negligible statistical fluctuations associated 
with finite particle numbers, which are difficult and expensive to filter 
out in such flow regimes \citep{pareschi2001,rjasanow2005}.
While several general purpose algorithms have been 
proposed, the most efficient techniques are problem 
specific~\citep{dimarco2014,cho2016,zhang2017,weinan2017}. 
These methods exploit the BTE's mathematical properties to arrive at an 
efficient algorithm, but are not generally applicable to other 
high-dimensional PDEs. 

We present a new algorithm based on 
tensor decompositions to solve the BTE in the Bhatnagar-Gross-Krook (BGK)
approximation~\citep{bhatnagar1954}. The algorithm relies 
on canonical tensor expansions \cite{boelens2018c}, 
combined with either alternating direction least squares methods \citep{reynolds2016,battaglino2018,acar2011,beylkin2009} or 
alternating direction Galerkin methods~\citep{douglas1971,venturi2018}
or any other version of the method of mean weighted 
residuals (MWR)~\citep{finlayson2013}. 
Unlike the BTE-specific numerical techniques, tensor-decomposition 
methods are general-purpose, i.e., they can be applied 
to other high-dimensional nonlinear PDEs \cite{DEKTOR2020,RODGERS2020},
including but not limited to the Hamilton-Jacobi-Bellman 
equation~\citep{akian2018}, the Fokker-Planck equation~\citep{risken1996}, and 
the Vlasov equation~\citep{hatch2012,dolgov2014,kormann2015}. 
This opens the possibility to use tensor methods in many research 
fields including chemical reaction networks in turbulent flows~\citep{monin2007b}, 
neuroscience~\citep{breakspear2017}, and approximation of functional 
differential equations~\citep{venturi2018}. Recently, we developed the tensor-decomposition 
method~\citep{boelens2018c} to solve a linearized BGK equation. In this paper, we extend 
it to the full BTE in the BGK approximation, i.e., 
to obviate the need for the assumption of small fluctuations and to 
allow for variable density, velocity, temperature, and collision frequency fields.

This paper is organized as follows. Section~\ref{sec:Boltzmann} contains a brief
overview of the Boltzmann-BGK equation. In section~\ref{sec:Numericaltensors}, we propose 
an efficient algorithm to compute its solution based canonical tensor expansions. 
Numerical experiments reported in section~\ref{sec:results} are used to demonstrate the algorithm's 
ability to accurately predict the equilibrium distribution of the system (stead-state), and the exponential relaxation 
to equilibrium (transient simulation) of non-equilibrium initial states. 
The algorithm exhibits $\mathcal{O} (N \log(N))$ scaling, 
where $N$ is the number of degrees of freedom in each of the 
phase variables. Main conclusions drawn from the numerical experimentation 
and future directions of research are summarized in section~\ref{sec:concl}.


\section{Boltzmann equation}
\label{sec:Boltzmann}

In the classical kinetic theory of rarefied gas dynamics, 
flow of gases is described in terms of a probability density 
function (PDF) $f(\mathbf x,\bm \xi ,t)$, which 
estimates the number of gas particles 
with velocity $\bm \xi\in\mathbb{R}^3$ at position  
$\mathbf x \in \mathbb{R}^3$ at time $t \in \mathbb R^+$, such that $\text dN = f \text d \mathbf x \text d \bm\xi$ with $N$ denoting the number of particles (in moles).  In the absence of external forces,
the PDF $f$ satisfies the Boltzmann equation~\citep{cercignani1988},
\begin{equation}
\frac{\partial f}{\partial t}+\bm \xi \cdot \nabla_{\mathbf x} f = Q(f,f), 
\label{BLZ}
\end{equation} 
where $Q(f,f)$ is the collision integral describing 
the effects of internal forces due to particle interactions. 
From the mathematical viewpoint, the collision integral is a 
functional of the PDF $f$, whose form depends on the 
microscopic dynamics. For example, in classical  
rarefied gas flows~\cite{cercignani1997,cercignani1994},
\begin{equation}
Q(f,f)(\mathbf x,\bm \xi,t)=\int_{\mathbb{R}^3}\int_{\mathbb{S}^2}
B(\bm \xi,\bm \xi_1,\bm \omega) \left| f(\mathbf x,\bm \xi',t)f(\mathbf x,\bm \xi_1',t)-
f(\mathbf x,\bm \xi,t)f(\mathbf x,\bm \xi_1,t) \right| \text d \bm \omega \text d\bm \xi_1.
\label{collisionOP}
\end{equation}
Here, $\bm \xi$ and $\bm \xi_1$ are the velocities of two particles before the collision; $\bm \xi' = (\bm \xi+\bm \xi_1+\left\|\bm \xi- \bm \xi_1 \right\|_2 \bm \omega)/2$ and  $\bm \xi_1' = (\bm \xi+\bm \xi_1-\left\|\bm \xi- \bm \xi_1 \right\|_2 \bm \omega)/2$ are these velocities after the collision;
 $\bm \omega$ is the unit vector to the three-dimensional unit sphere $\mathbb{S}^2$.
The collision kernel $B(\bm \xi ,\bm \xi_1,\bm \omega)$ is a non-negative function of the Euclidean 
2-norm $\left\|\bm \xi-\bm \xi_1\right\|_2$ and the scattering 
angle $\theta$ between the relative velocities before and 
after the collision,
\begin{equation}
\cos \theta = \frac{(\bm \xi - \bm \xi_1)\cdot \bm \omega}{\left\|\bm \xi -\bm \xi_1\right\|_2}.
\end{equation}
Specifically,
\begin{equation}
B(\bm \xi,\bm \xi_1,\bm \omega)=\left\| \bm \xi - \bm \xi_1 \right\|_2
\sigma(\left\| \bm \xi-\bm \xi_1\right\|_2\cos \theta), 
\end{equation}
where $\sigma$ is the cross-section scattering function \cite{cercignani1994}.
The collision operator \eqref{collisionOP} satisfies a system of three
conservation laws \citep{dimarco2014},
\begin{equation}
\int_{\mathbb{R}^3} Q(f,f) (\mathbf x,\bm \xi , t)\psi(\bm \xi) \text d \bm \xi=0, \qquad \text{$\psi(\bm \xi) = 1$ or $\bm \xi$ or $\left\|\bm \xi\right\|_2^2$},
\label{CL}
\end{equation}
for mass, momentum, and energy, respectively. It also satisfies the Boltzmann $H$-theorem,
\begin{equation}\label{eq:Htheorem}
\int_{\mathbb{R}^3} Q(f,f)(\mathbf x,\bm \xi, t)\log\left(f(\mathbf x,\bm \xi,t)\right)
\text d\bm \xi \leq 0,
\end{equation}
that implies that any equilibrium PDF, i.e.,
any PDF $f$ for which $Q(f,f) = 0$, is locally Maxwellian:
\begin{equation}
f_\text{eq}(\mathbf x,\bm \xi,t) = \frac{n(\mathbf x,t)}{(2\pi k_\text{B} T(\mathbf x,t) / m)^{3/2}}
\exp\left(-\frac{m \left\|\mathbf U(\mathbf x,t) - \bm \xi\right\|_2^2}{2 k_\text{B} T(\mathbf x,t)}\right).
\label{LocalMaxwell}
\end{equation}
Here $k_\text{B}$ is the Boltzmann constant; $m$ is the particle mass; and the number density $n$,  mean velocity $\mathbf U$, and temperature $T$ of a gas are defined as 
\begin{align}
n(\mathbf x,t) = &\int_{\mathbb{R}^3} f(\mathbf x,\bm \xi,t) \text d \bm \xi, \label{rho}\\ 
\mathbf U (\mathbf x,t) = &\frac{1}{n(\mathbf x,t)}\int_{\mathbb{R}^3}\bm \xi f(\mathbf x,\bm \xi,t) \text d \bm \xi,\label{u}\\ 
T(\mathbf x,t) = &\frac{1}{3n(\mathbf x,t)}\int_{\mathbb{R}^3} \left\|\mathbf U(\mathbf x,t) - \bm \xi \right\|_2^2 f(\mathbf x,\bm \xi,t) \text d \bm \xi \label{T}.
\end{align}
The Boltzmann equation \eqref{BLZ} is a nonlinear integro-differential equation in six dimensions plus time. By taking suitable averages over small volumes in position space, one can show that the Boltzmann equation is consistent with both the compressible Euler equations~\citep{Nishida1978,caflisch1980} and the 
Navier-Stokes equations \cite{Struchtrup2005,Levermore1996}.

\subsection{BGK approximation of the collision operator}
\label{sec:BoltzmannBGK}

The simplest collision operator satisfying the conservation laws~\eqref{CL} and the Boltzmann $H$-theorem~\eqref{eq:Htheorem} is the linear relaxation operator,
\begin{equation}
Q(f,f)=\nu(\mathbf x,t) \left[f_\text{eq}(\mathbf x,\bm \xi,t)-f(\mathbf x,\bm \xi,t)\right], \qquad \nu(\mathbf x,t)>0.
\label{BoBGK}
\end{equation}
It is known as the Bhatnagar-Gross-Krook (BGK) model \cite{bhatnagar1954}. The collision frequency $\nu(\mathbf x,t)$ is usually set to be proportional to the gas number-density and temperature~\citep{mieussens2000},
\begin{equation}
\nu(\mathbf x,t)=K n T^{1-\mu}.
\label{collfreq}
\end{equation}
The exponent of the viscosity law of the gas, $\mu$, depends on the molecular interaction potential and on 
the type of the gas; and $K=R_s T_\text{ref}/\mu_\text{ref} > 0$, where $\mu_\text{ref}$ is the gas viscosity at the reference temperature $T_\text{ref}$.   

The combination of~\eqref{BLZ} and~\eqref{BoBGK} yields the Boltzmann-BGK equation,
\begin{equation}
 \frac{\partial f(\mathbf x,\bm \xi,t)}{\partial t}+\bm \xi \cdot \nabla_{\mathbf x}  f(\mathbf x,\bm \xi,t) = \nu(\mathbf x,t) \left[f_\text{eq}(\mathbf x,\bm \xi,t)-f(\mathbf  x,\bm \xi,t)\right].
 \label{BOBGK}
\end{equation}
By virtue of~\eqref{rho}--\eqref{T}, both $f_\text{eq}(\mathbf x,\bm \xi,t)$ in~\eqref{LocalMaxwell} and $\nu(\mathbf x,t)$ in~\eqref{collfreq} are nonlinear functionals of the PDF $f(\mathbf x,\bm \xi,t)$. Therefore,  \eqref{BOBGK} is a nonlinear integro-differential PDE in 
six dimension plus time. It converges to the  Euler equations of incompressible fluid dynamics 
with the scaling $\mathbf x'=\epsilon \mathbf x$ and $t'=\epsilon t$,  in the 
limit $\epsilon\rightarrow 0$. However, it does not converge to the Navier-Stokes equations in this limit. 
Specifically, it predicts an unphysical Prandtl number~\cite{nassios2013}, which is 
larger than the one obtained with the full collision operator~\eqref{collisionOP}. 
The Navier-Stokes equations  can be recovered as $\epsilon$-limits 
of more sophisticated BGK models, e.g., the 
Gaussian-BGK model \cite{Andries2000}. 

In previous work \citep{boelens2018c}, we introduced additional 
simplifications, i.e., assumed  $\nu$ to be constant 
and the equilibrium density, temperature and velocity 
to be spatially homogeneous. If one additionally assumes 
$\mathbf U \equiv \mathbf 0$, the resulting model yields an equilibrium 
PDF $f_\text{eq}$ in~\eqref{LocalMaxwell}. The last assumption 
effectively decouples the BGK collision 
operator from the PDF $f(\mathbf x,\bm \xi,t)$. 
This, in turn, turns the BGK equation into a linear six-dimensional PDE.
In this paper, we develop a numerical method to solve 
the fully nonlinear Boltzmann-BGK equation.

\subsection{Scaling}
\label{sec:scaling}

Let us define the Boltzmann-BGK equation~\eqref{BOBGK} on 
a six-dimensional hypercube, such that $f : \Omega_{\mathbf x} \times \Omega_{\vec{\xi}} \times \mathbb R^+ \rightarrow \mathbb R^+$ with $\Omega_{\mathbf x}=[-b_{x},b_{x}]^{3}$ and $\Omega_{\vec{\xi}}=[-b_{\xi},b_{\xi}]^{3}$ representing the spatial domain and the velocity domain, respectively. Furthermore, we impose periodic boundary conditions on all the surfaces of this hypercube. 
We transform the hypercube $\Omega_{\mathbf x} \times \Omega_{\vec{\xi}}$ into the ``standard''
hypercube $\Omega_\pi = [-\pi,\pi]^{6}$, perform simulations 
in $\Omega_{\pi}$, and then map the numerical results back onto $\Omega_{\mathbf x} \times \Omega_{\vec{\xi}}$. This is accomplished by introducing dimensionless independent and dependent variables
\begin{equation*}
\tilde{\bm\xi} = \frac{\vec{\xi} \pi}{b_{\xi}}, \quad
\tilde{\mathbf x} = \frac{\mathbf x \pi}{b_{x}}, \quad 
\tilde t = \frac{t b_{\xi}}{b_{x}}, \quad
\tilde n = n b_x^3, \quad
\tilde{\mathbf U} = \frac{\mathbf U \pi}{b_{\xi}}, \quad
\tilde T = \frac{T}{T_{c}}, \quad
\tilde \nu = \frac{\nu \lambda}{b_{\xi}},
\end{equation*}
where $\lambda$ is the mean free path of a gas molecule, and $T_{c}$ is a characteristic temperature. Furthermore, we define the dimensionless Knudsen ($\text{Kn}$) and Boltzmann ($\text{Bo}$) numbers as
\begin{equation}
\txt{Kn} = \frac{\lambda}{b_x} \quad\text{and}\quad
\txt{Bo} = \frac{m b_\xi^2 }{\pi^2 k_\text{B} T_{c} }.
\label{KnBo}
\end{equation}
Then, the rescaled PDF $\tilde f(\tilde{\mathbf x},\tilde{\bm \xi},\tilde t) = f(\mathbf x,\bm \xi,t) b_x^3 b_\xi^3 / \pi^3$ satisfies a dimensionless form of the Boltzmann-BGK equation~\eqref{BOBGK},
\begin{equation}
 \frac{\partial \tilde f}{\partial \tilde t} = L(\tilde{\bm \xi}) \tilde f + C(\tilde{\mathbf x},\tilde{\bm \xi},\tilde t), 
 \qquad L(\tilde{\bm \xi}) \equiv - \tilde{\bm \xi} \cdot \nabla_{\tilde{\mathbf x}},
 \quad C(\tilde{\mathbf x},\tilde{\bm \xi},\tilde t) \equiv \frac{\tilde\nu}{\text{Kn}} (\tilde f_\text{eq} - \tilde f),
 \label{BOBGK_dimensionless}
\end{equation}
where
\begin{equation}
 \tilde{f}_{\txt{eq}} (\tilde{\mathbf x}, \tilde{\vec{\xi}}, \tilde t) =
  \frac{\tilde n }{ (2 \pi \tilde T / \text{Bo} )^{3/2}}
  \exp{\left(- \text{Bo} \frac{ \| \tilde{\vec{\xi}} - \tilde{\mathbf U} \|^2}
  {2 \tilde T } \right)},
\label{eqn:maxwellBoltzmann}  
\end{equation}
with
\begin{align}
 \tilde n(\tilde{\mathbf x}, \tilde t) =  & \int_{[-\pi,\pi]^3} \tilde f(\tilde{\mathbf x}, \tilde{\bm \xi}, \tilde t) \text d \tilde{\vec\xi},
\label{eqn:densityDim}\\
\tilde {\mathbf U} (\tilde{\mathbf x}, \tilde t) = & \, \frac{1}{ \tilde n(\tilde{\mathbf x}, \tilde t) } \int_{[-\pi,\pi]^3} 
 \tilde{\vec\xi} \tilde f(\tilde{\mathbf x}, \tilde{\bm\xi}, \tilde t) \text d \tilde{\vec\xi},
\label{eqn:velocityDim}\\ 
\tilde T(\tilde{\mathbf x}, \tilde t) = & \,
  \frac{\text{Bo}}{3 \tilde n(\tilde{\mathbf x}, \tilde t) } \int_{[-\pi,\pi]^3} \| \tilde{\vec\xi} - \tilde{\mathbf U}(\tilde{\mathbf x}, \tilde t) \|_2^2 \tilde f(\tilde{\mathbf x}, \tilde{\bm \xi}, \tilde t) \text d \tilde{\vec\xi}.
  \label{eqn:temperatureDim}  
\end{align}
For notational convenience, we drop the tilde below, while continuing to use the dimensionless quantities.

\section{A tensor method to solve the Boltzmann-BGK equation}
\label{sec:Numericaltensors}

Temporal discretization of the Boltzmann-BGK equation~\eqref{BOBGK_dimensionless} is complicated by the presence of the collision term $C(\mathbf x,\bm \xi,t)$, whose evaluation is computationally expensive. The combination of the Crank-Nicolson time integration scheme with alternating-direction least squares, implemented in~\citep{boelens2018c}, would require multiple evaluations of $C(\mathbf x,\bm \xi,t)$ per time step, undermining the efficiency of the resulting algorithm. To ameliorate this problem, we replace the Crank-Nicolson method with the Crank-Nicolson Leap Frog (CNLF) scheme~\citep{johansson1963,layton2012,kubacki2013,jiang2015}
\begin{equation}
  \frac{ f(\cdot, t_{n+1}) - f(\cdot, t_{n-1}) }{2 \Delta t}
= \frac{L(\vec{\xi}) f(\cdot, t_{n+1}) + L(\vec{\xi}) f(\cdot, t_{n-1})  }{  2} 
+ C(\cdot, t_n)+\tau_{n+1},
\label{CNLF}
\end{equation}
where $\tau_{n+1}$ is the local truncation error at time $t_{n+1}$.

The CNLF scheme has several advantages over other time-integration 
methods when applied to tensor discretization of the Boltzmann-BGK equation. 
First, being an implicit scheme, CNLF allows one to march forward in 
time by solving systems of linear equations on tensor manifolds 
with constant rank\footnote{Explicit time-integration algorithms 
require rank reduction \cite{grasedyck2010} as application of 
linear operators to tensors, tensor addition, and other tensor operations 
results in increased tensor ranks.}. Since such manifolds are {\em smooth}~\cite{breiding2018,uschmajew2013}, 
one can compute these solutions using, e.g., 
Riemannian quasi-Newton optimization~\citep{breiding2018,smith1994,yang2007} 
or alternating least squares~\citep{reynolds2016,uschmajew2012,bezdek2003}.
Second, CNLF facilitates the explicit calculation of the collision term $C \left(\vec{x},\vec{\xi}, t\right)$,
 and only {\em once} per time step.
To demonstrate this, we rewrite~\eqref{CNLF} as 
\begin{subequations}\label{timediscrete}
\begin{equation}
  \underbrace{
    [I - \Delta t \, L(\vec\xi) ]}_{A(\vec\xi)}
  f(\cdot,t_{n+1})
=
  \underbrace{
    [1 + \Delta t \, L(\vec\xi)] f(\cdot, t_{n-1})
  + 2 \Delta t \, C(\cdot,t_n)   
  }_{h(\mathbf x, \vec\xi, t_n, t_{n-1})}+2\Delta t \, \tau_{n+1},
\end{equation}
where $I$ is the identity operator; or
\begin{equation}
  A(\vec\xi) f(\cdot,t_{n+1}) = h(\mathbf x, \vec\xi, t_n, t_{n-1}) + 2\Delta t \, \tau_{n+1}.
\label{timediscrete1}
\end{equation}
\end{subequations}
Given $f(\mathbf x, \bm \xi, t_n)$ and  $f(\mathbf x, \bm \xi,t_{n-1})$,  
this equation allows us to compute $f(\mathbf x, \bm \xi,t_{n+1})$
by solving a linear system. In the numerical tensor setting described 
below, this involves only iterations in $f(\mathbf x, \bm \xi,t_{n+1})$, 
which makes it possible to pre-calculate the computationally expensive collision 
term $C(\mathbf x,\vec{\xi}, t)$ once per time step.

The choice of the time step $\Delta t$ in~\eqref{timediscrete} requires some care, since 
CNLF is conditionally stable~\cite{Hurl2014}. We transform this scheme into an unconditionally stable one 
by using, e.g., the Robert-Asselin-Williams (RAW) filter~\citep{kwizak1971,williams2009,williams2011}.

\subsection{Canonical tensor decomposition and alternating least squares (ALS)} 
\label{sec:als}

We expand the PDF $f(\mathbf x, \vec{\xi}, t_n)$ in a 
truncated canonical tensor series \cite{boelens2018c,beylkin2009},
\begin{equation}
  f(\mathbf x,\bm \xi,t_n)\simeq
  \sum_{l=1}^{r_{l}} 
  {f}_{1}^{l}(x_1,t_n)
  {f}_{2}^{l}(x_2,t_n)
  {f}_{3}^{l}(x_3,t_n)
  {f}_{4}^{l}(\xi_1,t_n)
  {f}_{5}^{l}(\xi_2,t_n) 
  {f}_{6}^{l}(\xi_3,t_n).
\label{expF}
\end{equation}
The separation rank $r_{l}$ is chosen adaptively to keep the 
norm of the residual below a pre-selected threshold 
at each time $t_n$. To simplify the notation, we introduce the combined 
position-velocity vector $\bm \zeta = (\mathbf x,\bm \xi)$
and rewrite \eqref{expF} as 
\begin{equation}
 f(\bm \zeta, t_n) \simeq
 \sum_{l=1}^{r_{l}}\prod_{k=1}^6 f_k^l(\zeta_k, t_n),
\qquad  \bm \zeta\in \Omega_{\pi}=[-\pi,\pi]^6.
\label{expF1}
\end{equation}  
Next, we expand each function ${f}_{k}^{l} \brc{\zeta_{k}, t_{n}}$ in a finite-dimensional Fourier basis 
$\phi_s(\zeta_k)$~\cite{Hesthaven2007} on $[-\pi,\pi]$,
\begin{equation}
f_k^l(\zeta_k, t_n) =
\sum_{s=1}^Q \beta_{k,s}^{l}(t_n) \phi_s(\zeta_k), 
\label{coeff}
\end{equation}
where $Q$ is the number of modes of the Fourier-series expansion, $\beta_{k,s}^{l} \brc{t_{n}}$ are the (unknown)
Fourier coefficients, and $\phi_s(\zeta_k)$ are the orthogonal trigonometric functions. 
Substituting \eqref{expF1} into \eqref{timediscrete} yields the residual 
\begin{equation}
 R(\vec{\zeta}, t_{n+1}, t_{n}, t_{n-1}) = \sum_{l=1}^{r_{l}} A \brc{\vec{\xi}} 
  {f}_{1}^{l} \brc{\zeta_{1}, t_{n+1}} \ldots 
  {f}_{6}^{l} \brc{\zeta_{6}, t_{n+1}}
- h \brc{\vec{\zeta}, t_{n}, t_{n-1}}.
  \label{res}
\end{equation}
The coefficients $\beta_{k,s}^l(t_n)$ are obtained 
by minimizing the $L^2$ norm of this residual with respect to 
\begin{equation}
\bm \beta(t_{n+1}) = [\vec{\beta}_{1}(t_{n+1}), 
\ldots, \vec{\beta}_{6}(t_{n+1})].
\label{beta}
\end{equation}
The $k$th vector $\vec{\beta}_k(t_{n+1}) =
  [
    ( \beta_{k,1}^1, \ldots \beta_{k,Q}^1),
    \ldots
    (\beta_{k,1}^r,\ldots \beta_{k,Q}^r )
  ]^\top$, for $k=1,\ldots,6$,
collects the degrees of freedom representing the PDF $f$ in \eqref{expF1} corresponding to the phase variable $\zeta_k$ at time $t_{n+1}$, in accordance with~\eqref{coeff}.

We employ the alternating least squares (ALS) algorithm \citep{bezdek2003, ortega1970} to solve the minimization problem
\begin{equation}
\min_{\vec{\beta}_{k}} 
\left\|R \brc{\vec{\zeta}, t_{n+1}, t_{n}, t_{n-1}}\right\|^{2}_{L^2(\Omega_\pi)} 
\label{ALS0}
\end{equation}
sequentially and iteratively for $k=1,\dots, 6$. The ALS algorithm is locally equivalent to 
the linear block Gauss--Seidel iteration method applied to the Hessian 
of the residual $R$. As a consequence, it converges linearly 
with the iteration number \citep{uschmajew2012}, 
provided that the Hessian is positive definite 
(except on a trivial null space associated with the scaling 
non-uniqueness of the canonical tensor decomposition).
Each minimization in \eqref{ALS0} yields an  
Euler-Lagrange equation, 
\begin{equation}
  \mathbf M_{k}(t_{n+1}) \vec{\beta}_{k}(t_{n+1})
= \bm \gamma_{k}(t_{n+1}), \qquad k=1,\dots, 6.
\label{eqn:bte}
\end{equation}
Its expanded form reads
{\fontsize{8pt}{8pt}\selectfont{
\begin{equation}
  \underbrace{\sqrbrc{
    \!\!\!\!
    \vphantom{\rule{0pt}{3.1cm}}
    \begin{array}{c@{}c@{}c@{}c}
    \sqrbrc{
    \vphantom{\rule{0pt}{0.95cm}}
      \begin{array}{c@{}c@{}c}
         (M_k)_{1,1}^{1,1} & \cdots &  (M_k)_{Q,1}^{1,1} \\
        \vdots        & \ddots & \vdots        \\
         (M_k)_{1,Q}^{1,1} & \cdots &  (M_k)_{Q,Q}^{1,1} \\
      \end{array}
    } &
    \sqrbrc{
    \vphantom{\rule{0pt}{0.95cm}}
      \begin{array}{c@{}c@{}c}
         (M_k)_{1,1}^{2,1} & \cdots &  (M_k)_{Q,1}^{2,1} \\
        \vdots        & \ddots & \vdots        \\
         (M_k)_{1,Q}^{2,1} & \cdots &  (M_k)_{Q,Q}^{2,1} 
      \end{array}
    } &
    \cdots &
    \sqrbrc{
    \vphantom{\rule{0pt}{0.95cm}}
      \begin{array}{c@{}c@{}c}
         (M_k)_{1,1}^{r,1} & \cdots &  (M_k)_{Q,1}^{r,1} \\
        \vdots        & \ddots & \vdots        \\
         (M_k)_{1,Q}^{r,1} & \cdots &  (M_k)_{Q,Q}^{r,1} 
      \end{array}
    } \\
    \sqrbrc{
    \vphantom{\rule{0pt}{0.95cm}}
      \begin{array}{c@{}c@{}c}
         (M_k)_{1,1}^{1,2} & \cdots &  (M_k)_{Q,1}^{1,2} \\
        \vdots        & \ddots & \vdots        \\
         (M_k)_{1,Q}^{1,2} & \cdots &  (M_k)_{Q,Q}^{1,2} 
      \end{array}
    } &
    \sqrbrc{
    \vphantom{\rule{0pt}{0.95cm}}
      \begin{array}{c@{}c@{}c}
         (M_k)_{1,1}^{2,2} & \cdots &  (M_k)_{Q,1}^{2,2} \\
        \vdots        & \ddots & \vdots        \\
         (M_k)_{1,Q}^{2,2} & \cdots &  (M_k)_{Q,Q}^{2,2}
      \end{array}
    } &
    \cdots &
    \sqrbrc{
    \vphantom{\rule{0pt}{0.95cm}}
      \begin{array}{c@{}c@{}c}
         (M_k)_{1,1}^{r,2} & \cdots &  (M_k)_{Q,1}^{r,2} \\
        \vdots        & \ddots & \vdots        \\
         (M_k)_{1,Q}^{r,2} & \cdots &  (M_k)_{Q,Q}^{r,2}
      \end{array}
    } \\
    \vdots &
           &
    \ddots &
    \vdots \\
    \sqrbrc{
    \vphantom{\sqrbrc{\rule{0pt}{0.95cm}}}
      \begin{array}{c@{}c@{}c}
         (M_k)_{1,1}^{1,r} & \cdots &  (M_k)_{Q,1}^{1,r} \\
        \vdots        & \ddots & \vdots        \\
         (M_k)_{1,Q}^{1,r} & \cdots &  (M_k)_{Q,Q}^{1,r} 
      \end{array}
    } &
    \sqrbrc{
    \vphantom{\sqrbrc{\rule{0pt}{0.95cm}}}
      \begin{array}{c@{}c@{}c}
         (M_k)_{1,1}^{2,r} & \cdots &  (M_k)_{Q,1}^{2,r} \\
        \vdots        & \ddots & \vdots        \\
         (M_k)_{1,Q}^{2,r} & \cdots &  (M_k)_{Q,Q}^{2,r} 
      \end{array}
    } &
    \cdots &
    \sqrbrc{
    \vphantom{\sqrbrc{\rule{0pt}{0.95cm}}}
      \begin{array}{c@{}c@{}c}
         (M_k)_{1,1}^{r,r} & \cdots &  (M_k)_{Q,1}^{r,r} \\
        \vdots        & \ddots & \vdots        \\
         (M_k)_{1,Q}^{r,r} & \cdots &  (M_k)_{Q,Q}^{r,r} 
      \end{array}
    } \\
    \end{array}
    \!\!\!\!
  }}_{\mathbf M_k}
 \underbrace{\sqrbrc{
    \!\!\!\!
    \vphantom{\rule{0pt}{3.1cm}}
    \begin{array}{c}
    \sqrbrc{
      \vphantom{\rule{0pt}{0.95cm}}
      \begin{array}{c}
        \beta_{k,1}^{1} \\
        \vdots \\
        \beta_{k,Q}^{1}
      \end{array}
    } \\
    \sqrbrc{
      \vphantom{\rule{0pt}{0.95cm}}
      \begin{array}{c}
        \beta_{k,1}^{2} \\
        \vdots \\
        \beta_{k,Q}^{2}
      \end{array}
    } \\
    \vdots \\
    \sqrbrc{
      \vphantom{\rule{0pt}{0.95cm}}
      \begin{array}{c}
        \beta_{k,1}^{r} \\
        \vdots \\
        \beta_{k,Q}^{r}
      \end{array}
    }
    \end{array}
    \!\!\!\!
  }}_{\bm \beta_k}
=
  \underbrace{\sqrbrc{
    \!\!\!\!
    \vphantom{\rule{0pt}{3.1cm}}
    \begin{array}{c}
    \sqrbrc{
      \vphantom{\rule{0pt}{0.95cm}}
      \begin{array}{c}
        \gamma_{k,1}^{1} \\
        \vdots \\
        \gamma_{k,Q}^{1}
      \end{array}
    } \\
    \sqrbrc{
      \vphantom{\rule{0pt}{0.95cm}}
      \begin{array}{c}
        \gamma_{k,1}^{2} \\
        \vdots \\
        \gamma_{k,Q}^{2}
      \end{array}
    } \\
    \vdots \\
    \sqrbrc{
      \vphantom{\rule{0pt}{0.95cm}}
      \begin{array}{c}
        \gamma_{k,1}^{r} \\
        \vdots \\
        \gamma_{k,Q}^{r}
      \end{array}
    }
    \end{array}
    \!\!\!\!
  }}_{\bm \gamma_k}
\nonumber  
\end{equation}
}}
where 
\begin{equation}
\brc{M_{k}}_{s,q}^{l,z}(t_{n+1})=\int_{\Omega_\pi} 
\Big[ A(\vec\xi) \phi_s(\zeta_k) 
\prod_{\substack{j=1\\j\neq k}}^6 f_j^l(\zeta_j, t_{n+1})
\Big] \Big[
A(\vec{\xi}) \phi_q(\zeta_k) 
\prod_{\substack{j=1\\j\neq k}}^6 f_j^z(\zeta_j,t_{n+1})
\Big] \text d \vec{\zeta}.
  \label{eqn:double}
\end{equation}
Since the linear operator, first defined in~\eqref{timediscrete}, is fully separable (with rank 4), the 6D integral in \eqref{eqn:double} turns into the sum of the product of 1D integrals. The right hand side of \eqref{eqn:double} is
\begin{equation}
\begin{aligned}
\gamma_{k,q}^{l}= &\int_{\Omega_\pi}
  h(\vec\zeta, t_n, t_{n-1}) A(\vec\xi) \phi_q(\zeta_k) 
\prod_{\substack{j=1\\j\neq k}}^6 
f_j^l(\zeta_j,t_{n+1}) \text d \vec{\zeta}
 \\
=& \sum_{m=1}^{r_m} \int_{\Omega_\pi} \Big\{ [I + \Delta t L(\bm \xi) ]
\prod_{\substack{j=1}}^6 f_j^m( \zeta_j, t_{n-1}) \Big\}
\Big\{  A(\vec\xi) \phi_q(\zeta_k)
 \prod_{\substack{j=1\\j\neq k}}^6 f_j^l(\zeta_j, t_{n+1}) \Big\}
\text d\bm \zeta 
\\
& + 2 \Delta t \int_{\Omega_\pi}
  C(\vec\zeta, t_n) A(\vec\xi) \phi_q(\zeta_k) 
 \prod_{\substack{j=1\\j\neq k}}^6 f_j^l(\zeta_j, t_{n+1}) 
\text d \vec\zeta.
  \label{eqn:gamma} 
\end{aligned}
\end{equation}
The 6D integrals under the sum are, as before, a sum of the products of 1D integrals, because $L$ is separable with rank 3.

\subsection{Evaluation of the BGK collision term}

The BGK collision term $C(\vec\zeta, t_n)$ in \eqref{eqn:gamma},
\begin{align}
  C(\vec{\zeta},t_n) =
  \frac{\nu(\mathbf{x}, t_n) }{\text{Kn}} \sqrbrc{f_{\txt{eq}}(\mathbf x,\bm \xi,t_n)
  - f(\mathbf x, \vec\xi, t_n) },
\label{eqn:c}
\end{align}
is evaluated by using a canonical tensor decomposition. The equilibrium distribution $f_{\txt{eq}}$, defined in~\eqref{eqn:maxwellBoltzmann}, is the product of one 3D function and three 4D functions,
\begin{align}
f_{\text{eq}}(\mathbf x,\bm \xi, t_n)=
  \frac{n(\vec{x}, t_n) }{ [2 \pi T(\mathbf x, t_n) / \text{Bo} ]^{3/2}} \prod_{m=1}^3
  \exp\left( - \text{Bo} \frac{ [ \xi_m - U_m(\mathbf x, t_n) ]^2 }{ 2 T(\mathbf x, t_n) }  \right).
\label{Eq}
\end{align}
Each of these terms are expanded in a canonical tensor series once the number density, 
velocity and temperature are computed using \eqref{eqn:densityDim}--\eqref{eqn:temperatureDim}. 
The integrals in these expressions are reduced to the products of 1D integrals, once 
the canonical tensor expansion \eqref{expF1} is available.
To calculate the normalization by $n(\mathbf x, t_n)$ in~\eqref{eqn:densityDim}--\eqref{eqn:temperatureDim} and the normalization by $T(\mathbf x, t_n)$ in~\eqref{Eq}, we employ a Fourier collocation method with $N$ points in each variable. That turns all the integrals into Riemannian sums, 
and the functions $f_k^l(\zeta_k,t_n)$ into
\begin{align}
f_k^l(\zeta_{k,j}, t_n) = \frac{1}{4 \pi} \left[
    \sum_{s = -N/2+1}^{N/2} \beta_{k,s}^{l} \brc{t_{n}} \text{e}^{i s \zeta_{k,j} } + \sum_{s = -N/2}^{N/2-1} \beta_{k,s}^l(t_n) \text{e}^{i s \zeta_{k,j} } \right]
\label{eqn:transform}
\end{align}
with inverse
\begin{equation}
  \beta_{k,s}^{l} \brc{t_{n}}
=
  h \sum_{j=1}^{N} f_k^l(\zeta_{k,j}, t_n) \text{e}^{-i s \zeta_{k,j} }.
\label{inverse}
\end{equation}
Here, $h = 2 \pi / N$ and $\zeta_{k,j}=-\pi+jh$ with $j=1,\dots,N$.
The form of~\eqref{eqn:transform} with two
different summation intervals is chosen to ensure that the highest wave-number is
treated symmetrically~\cite{trefethen2000}. The forward and backward transform
is performed efficiently with the Fast Fourier Transform (FFT) 
and its inverse. The collision frequency 
 $\nu(\mathbf x,t)= K n(\mathbf x,t)T(\mathbf x,t)^{1-\mu}$ 
is also represented by a canonical tensor series, once $n(\mathbf x,t)$ 
and $T(\mathbf x,t)$ are available. To speed up the tensor decomposition 
algorithm, the result from the previous time step is used as the initial 
guess for the new decomposition.

\subsection{Parallel ALS algorithm}

Our implementation of the parallel ALS algorithm for solving the Boltzmann transport equation is encapsulated Algorithm~\ref{alg:als}.\footnote{Details of the algorithmic implementation of our method and description of the various subroutines mentioned herein are provided in the Supplemental Material.} The subroutine
\textproc{Initialization} initializes all the variables needed to run the code. 
This includes, initialization of the time-step size $\Delta t$, the number of time
steps $n_{\text{max}}$, the number of collocation points, and the initialization
of the coefficients $\beta_{\txt{New}}$, $\beta_{\txt{Now}}$, and
$\beta_{\txt{Old}}$. These coefficients store, respectively, the values of $f\vec{\zeta},t_{n+1})$, $f(\vec{\zeta},t_{n})$, and $f(\vec{\zeta},t_{n-1})$ in Fourier space. In addition, the operators $A^{+}$, $A^{-}$, and
$\txt{Exp}_{{A}^{+}}$ are initialized. The operators $A^{+}$ and $A^{-}$ represent $1 \pm \Delta t \, L(\vec{\xi})$, respectively. The rank separated operator
$\txt{Exp}_{{A}^{+}}$ represents operator $A^{+}$ acting on the Fourier basis
function $\phi_{s}$. The representation of $A^{+}$ as
a single operator reduces the loss of accuracy that comes with
multiplying different operators.

\begin{algorithm}[t!]
\caption{Parallel ALS algorithm}
\label{alg:als}
\begin{algorithmic}[1]
\Procedure{Main}{}
\State{\Call{Initialization}{}} \Comment{Load variables and allocate matrices}
\State
\For{$n \gets 1:n_{\txt{max}}$}
\State{$C \gets \Call{computeArrayC}{\beta_{\txt{Now}}}$} \Comment{Compute collision operator}
\State
\State{$\beta_{\txt{New}} \gets \Call{randBeta}{\beta_{\txt{New}}}$} \Comment{Add some random noise}
\State{$\epsilon_{| \beta |} \gets \epsilon_{\txt{Tol}} + 10^{6}$} \Comment{Reset stop criterion}
\While{$\epsilon_{| \beta |} > \epsilon_{\txt{Tol}}$}
\State{$\beta_{\txt{Int}} \gets \beta_{\txt{New}}$} \Comment{Set intermediate value of beta}
\State
\State{$N \gets \Call{computeArrayN}{A^{+},A^{-},\txt{Exp}_{A^{+}},\beta_{\txt{Old}},\beta_{\txt{New}}}$}
\State{$O \gets \Call{computeArrayO}{C    ,A^{+},\txt{Exp}_{A^{+}},\beta_{\txt{New}}}$}
\State{$\gamma \gets N + 2 \; \Delta t \; O$}
\State
\ParFor{$d \gets 1:6$} \Comment{Iterate over dimensions}
\State{$M \brc{d} \gets \Call{computeArrayM}{A^{+},\txt{Exp}_{A^{+}} \brc{d},\beta_{\txt{New}},d}$}
\EndParFor
\State
%
\ParFor{$d \gets 1:6$}
\State{$\beta_{\txt{New}} \brc{d} \gets \Call{computeBetaNew}{\beta_{\txt{New}} \brc{d}, M \brc{d}, \gamma \brc{d} }$}
\EndParFor
\State
\State{$\beta_{\txt{New}} \gets \Call{kreal}{\beta_{\txt{New}}}$} \Comment{Keep only real part of solution}
\State{$\epsilon_{| \beta |} \gets \Call{computeNormBeta}{\beta_{\txt{Int}},\beta_{\txt{New}}}$} \Comment{Check for convergence}
\State
\If{$ \epsilon_{| \beta |} > \epsilon_{\txt{Tol}}$}
\State{$\beta_{\txt{New}} \gets \Call{computeBetaNewDelta}{\beta_{\txt{New}},\beta_{\txt{Int}}}$} \Comment{Update}
\Else
\State{$\beta_{\txt{New}},\beta_{\txt{Now}} \gets \Call{computeRAW}{\beta_{\txt{New}},\beta_{\txt{Now}},\beta_{\txt{Old}}}$} \Comment{Apply RAW filter}
\EndIf
\State
\EndWhile
\State{$\beta_{\txt{Old}} = \beta_{\txt{Now}}$} \Comment{Update for next time step}
\State{$\beta_{\txt{Now}} = \beta_{\txt{New}}$}
\EndFor
\EndProcedure
\end{algorithmic}
\end{algorithm}

For every time step, \textproc{computeArrayC} evaluates the BGK collision operator
$C(\vec\zeta,t_n)$ in~\eqref{eqn:c}, and \textproc{randBeta} adds a small amount of noise to $\beta_{\txt{New}}$ to
prevent the algorithm from getting stuck in a local minimum. The \textproc{while}
loop contains the parallel ALS algorithm. The functions \textproc{computeArrayN}
and \textproc{computeArrayO} inside the loop compute the two different contributions to the
vector $\gamma$ in~\eqref{eqn:gamma}. The function
\textproc{computeArrayM} computes the array $M$ inside a parallel \textproc{for} loop,
which completes the set of equations~\eqref{eqn:bte}.

The least-squares routine inside the function \textproc{computeBetaNew} updates $\beta_{\txt{New}}$ every iteration inside a parallel \textproc{for} loop. Because
of the large number of degrees of freedom in the above system, there are multiple
(local minimum) solutions, which are not necessarily real and mass conserving.
We ameliorate this problem by adding a constraint that, for every
iteration, only the real part of the solution for $\beta_{\txt{New}}$ is kept. This calculation is
performed by the function \textproc{kreal}. The function
\textproc{computeNormBeta} then computes the convergence criterion, $\epsilon_{|
\beta |}$, by comparing the current value of $\beta_{\txt{New}}$ to its value at
the end of the previous iteration, $\beta_{\txt{Int}}$. If convergence has not
been reached, \textproc{computeBetaNewDelta} updates $\beta_{\txt{New}}$ for the
next iteration. If convergence has been reached, \textproc{computeRAW} applies
the RAW filter \citep{kwizak1971,williams2009,williams2011} to make the CNLF
algorithm unconditionally stable, and the algorithm moves on to the next time step.

\section{Numerical results}
\label{sec:results}

\indent
In this section we study the accuracy and computational 
efficiency of the proposed ALS-CNLF tensor method to solve the 
Boltzman-BGK equation. (Unless specified otherwise, values of the simulation parameters used in our numerical experiments are collated in 
Table~\ref{tbl:param}.) We start by validating our solver on a prototype problem 
involving the Boltzmann-BGK equation in one spatial 
dimension. Then we consider simulations of the full 
Boltzmann-BGK equation in three spatial dimensions. 
The 3D numerical results are split up into steady-state 
transient problems. The steady-state problem is used to validate the 
code against an analytical equilibrium solution to 
the Boltzmann transport equation. It also enables us to investigate 
 the error convergence as function of various parameters, 
and the code performance as function of the number of 
collocation points and the number of processor 
cores.  The transient problem
starts with an initial distribution away from equilibrium, including a non-zero
average velocity in the $x_{1}$ direction, and looks at temporal evolution of
the PDF $f$. We compare our code with a spectral code in 2D and present the results for the full 6D Boltzmann-BGK equation.

{\renewcommand{\arraystretch}{1.1}
\begin{table}[htbp]
\centering
\begin{tabular}{lll}
\hline
\bf{Variable} & \bf{Value}                   & \bf{Description} \\
\hline                                         
$\Delta t$    & $0.025$                      & Dimensionless time step \\ 
$K$      & $1.0$                        & Collision frequency pre-factor \\
$\mu$         & $0.5$                        & Collision frequency temperature exponent \\
$\txt{Kn}$    & $1$                          & Knudsen number \\
$\txt{Bo}$    & $3.65$                       & Boltzmann number \\
$\epsilon_{\txt{Tol}}$ & $5.0 \cdot 10^{-5}$ & Tolerance on ALS iterations \\
\hline
\end{tabular}
\caption{Values of parameters used in the simulations of the Boltzmann-BGK equation in three spatial dimensions.}
\label{tbl:param}
\end{table}

\subsection{Transient dynamics in one spatial dimension}
\label{sec:1Dcomparison}

To validate the proposed Boltzmann-BGK tensor solver, 
we first compute the numerical solution of \eqref{BOBGK_dimensionless} 
in one spatial dimension, and compare it with an accurate benchmark 
solution obtained with the high-order Fourier pseudo-spectral 
method~\cite{Hesthaven2007}. Specifically, we study the 
 initial-value problem
\begin{equation}
\frac{\partial f(x,\xi,t)}{\partial t}+\xi\frac{\partial f(x,\xi,t)}{\partial x} = \frac{KT(x,t)^{1-\mu}}{\text{Kn}} \left[f_\text{eq}(x,\xi,t)-f(x,\xi,t)\right],
\label{1DBGK}
\end{equation}
\begin{equation}
  f \brc{x, \xi, 0}
= \frac{n_{0}}{\sqrt{(2 \pi T_{0} / \txt{Bo})}} 
\exp{\brc{-\frac{\txt{Bo}}{2 T_{0}} (U_{0} - \xi)^{2}}},
\label{1DIC}
\end{equation}
with
\begin{align}
n_0 = 1.0 + 0.3 \cos{(2 x)}, \qquad
U_0 = 1.0 + 0.1 \sin{(3 x)}, \qquad
T_0 = 1.0.
\label{TIC}
\end{align}
The benchmark numerical solution is 
constructed by solving~\eqref{1DBGK}--\eqref{TIC} 
in the periodic box $(x,\xi)\in[-\pi,\pi]^2$ with a 
Fourier pseudo-spectral method~\cite{Hesthaven2007} (odd expansion 
on a $61\times 61$ grid) and an explicit two-steps 
Adams-Bashforth temporal integrator 
with $\Delta t= 0.0005$. The remaining 
parameters are set to the values from table~\ref{tbl:param}, except 
$\txt{Kn}=10$. Figure~\ref{fig:contours} exhibits
contour plots of the benchmark PDF solution at 
three times. 

\begin{figure}[htbp]
\center
\includegraphics[width=\textwidth]{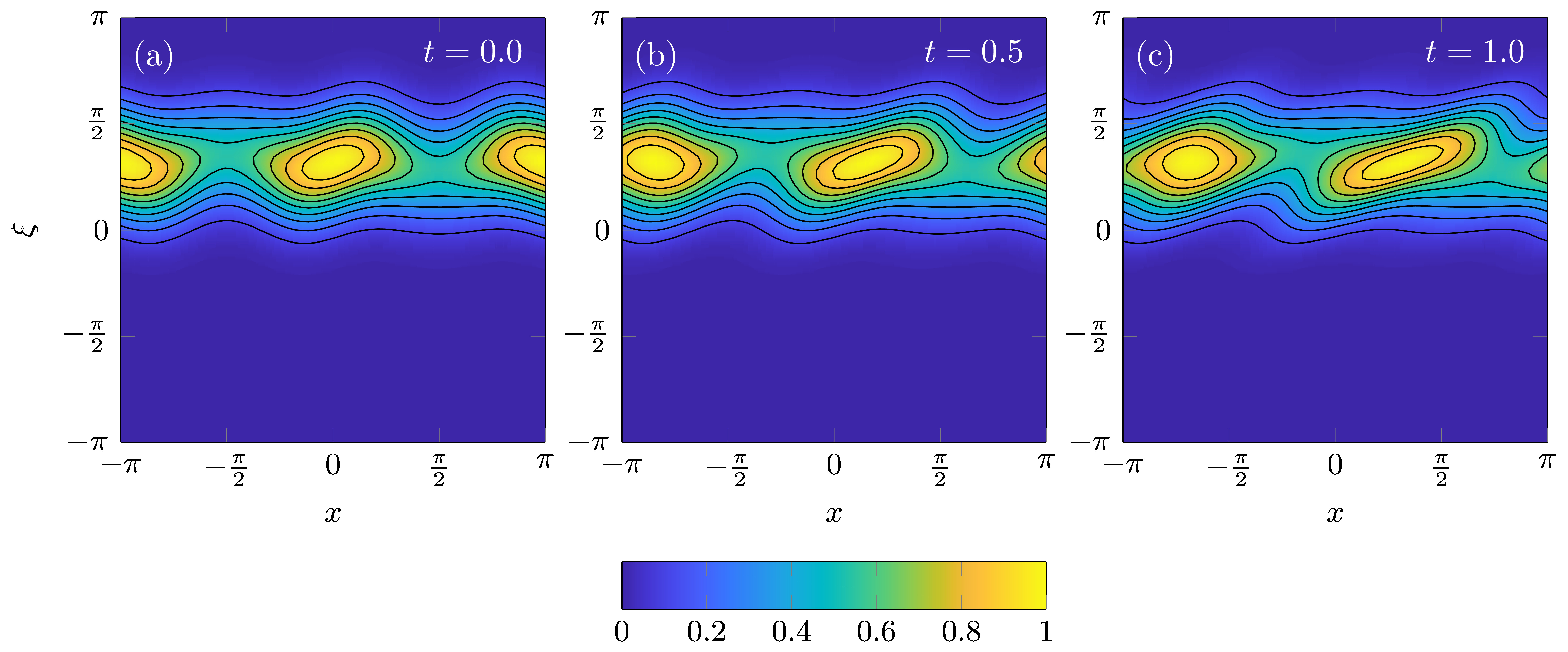}
\caption{Contour plots of the reference solution to the 1D Boltzmann-BGK 
initial-value problem~\eqref{1DBGK}--\eqref{TIC}, the PDF $f_\text{spc}(x,\xi,t)$, computed with 
the high-order Fourier pseudo-spectral method and explicit two-steps Adams-Bashforth temporal integrator.  }
\label{fig:contours}
\end{figure}

We report the accuracy of our ALS-CNLF tensor algorithm\footnote{
The tolerance for the ALS iterations is set to  
$\epsilon_{\txt{Tol}}= 10^{-12}$, while 
the other simulation parameters, such as $\Delta t$ 
and the number of grid points are the same as 
in the Fourier pseudo-spectral method.} 
by comparing its prediction, $f_\text{ALS}$, with the reference solution, $f_\text{spc}$, in terms of two metrics. The first is the $L^2$ norm 
\begin{equation}
\left\| f_{\txt{ALS}}(x,\xi,t) -  f_{\txt{spc}}(x,\xi,t) \right\|
= \left( \int_{[-\pi,\pi]^{2}}
 \left[f_{\txt{ALS}}(x,\xi,t) - f_{\txt{spc}}(x,\xi,t)\right]^2
\text dx \text d\xi \right)^{\!\! 1/2}.
\label{L2}
\end{equation}
The second is the Kullback-Leibler divergence 
\begin{equation}
  D_{\txt{KL}} (F_{\txt{ALS}} || F_{\txt{spc}}) 
= \int_{[-\pi,\pi]^{2}}
f_{\txt{ALS}}(x,\xi,t)
\log \left[\frac{ f_{\txt{ALS}}(x,\xi,t) }{ f_{\txt{spc}} 
(x,\xi,t) } \right] 
\text dx \text d\xi.
\label{KL}
\end{equation}
In figure~\ref{fig:l2kl}, we plot these two metrics as function of time $t$
and the tensor rank $r$, which is kept constant throughout 
the simulation. Bot the $L^2$ error and 
the Kullback-Leibler divergence decrease with the tensor rank $r$.

\begin{figure}[htbp]
\center
\includegraphics[width=0.8\textwidth]{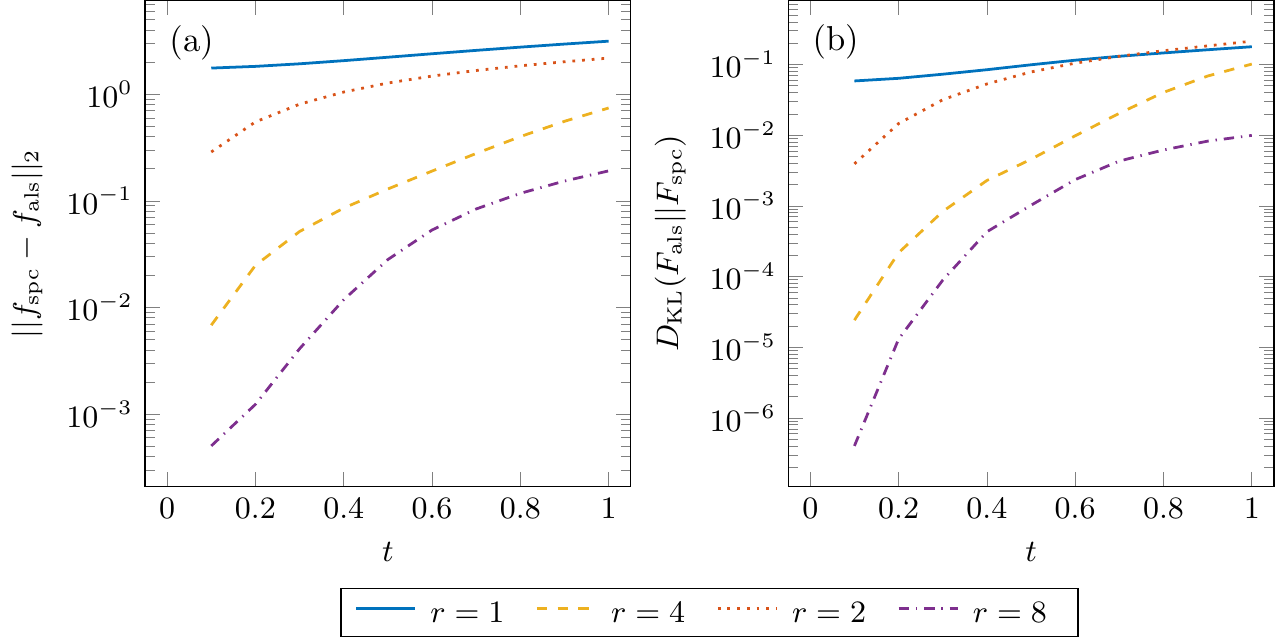}
\caption{Boltzmann-BGK problem \eqref{1DBGK}--\eqref{TIC} in one spatial dimension: $L^2$ error and Kullback-Leibler divergence of the 
ALS-CNLF tensor solution, $f_{\txt{ALS}}$, relative to the reference solution, $f_{\txt{spc}}$, 
versus time. Both the $L^2$ error and the Kullback-Leibler divergence 
decrease with the tensor rank $r$.}
\label{fig:l2kl}
\end{figure}

}

\subsection{Steady-state simulations in three spatial dimensions}

\begin{figure}[htbp]
\center
\includegraphics[width=\textwidth]{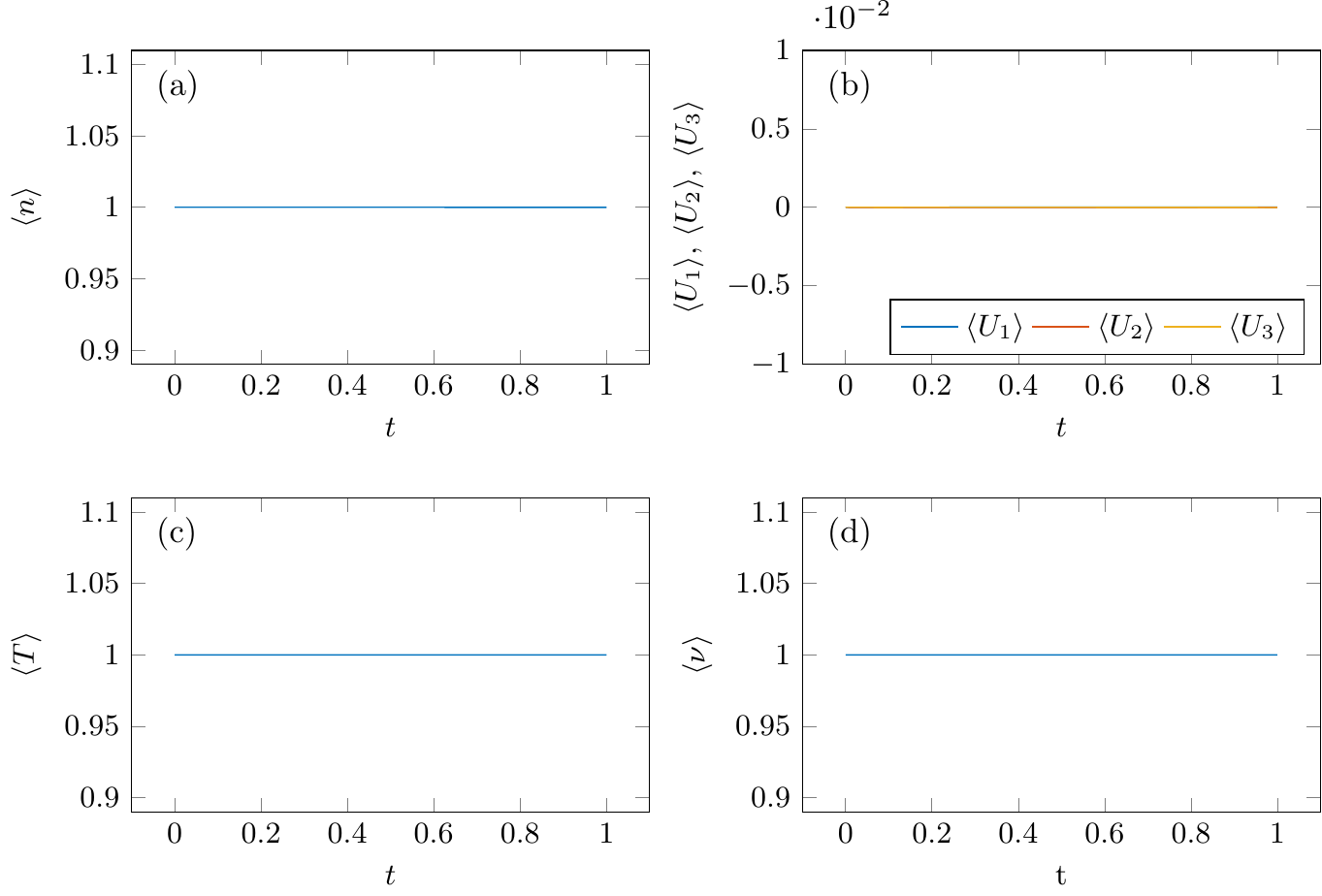}
\caption{Temporal variability of relative errors in the spatial averages of density, $\avr{n}$,
velocity components, $\avr{U_{1}}$, $\avr{U_{2}}$, and $\avr{U_{3}}$, and collision frequency, $\avr{\nu}$, as well as of the spatially averaged temperature, $\avr{T}$. The number of collocation points per dimension is $N=64$. The results show that the code is mass conserving, i.e., the moments of the PDF $f$ at equilibrium are constant constant in time up to a small ($10^{-5}$) error.}
\label{fig:fieldsAvrConst}
\end{figure}

Since convergence of the ALS algorithm is not guaranteed, e.g.,
\cite{uschmajew2012,comon2009}, and since the equilibrium distribution is one of
the few analytical solutions to the Boltzmann transport equation in six dimensions,
we report the code's behavior at equilibrium. The initial condition in this experiment is the Maxwell-Boltzmann equilibrium PDF, whose moments, as defined in~\eqref{eqn:densityDim}--\eqref{eqn:temperatureDim}, are set to $n(\mathbf x,0) = 1$, $U_1(\mathbf x,0) = U_2(\mathbf x,0) = U_3(\mathbf x,0) = 0$, and $T (\mathbf x,0) =
1$. The simulation was ran till $t = 1$. Figure~\ref{fig:fieldsAvrConst} shows temporal variability of the relative errors of the spatial averages, $\avr{n}$ and $\avr{T}$, of the moments $n$ and $T$ and that of the collision frequency, $\avr{\nu}$. Since the initial values of the velocities, $\avr{U_{1}}$,
$\avr{U_{2}}$, and $\avr{U_{3}}$, are zero, figure~\ref{fig:fieldsAvrConst}b
 shows only their average values as function of time. For any quantity $a (\mathbf x, t) $, the spatial average is defined as
\begin{equation}
  \avr{a (t)}
=   \frac{1}{\brc{2 \pi}^3} \int_{[-\pi,\pi]^{3}} a (\mathbf x,t) \text d \mathbf x.
\end{equation}
As expected, all the spatially averaged moments at equilibrium remain approximately constant
with time, with small (on the order of $10^{-5}$) deviations representing the numerical error.

\begin{figure}[htbp]
\center
\includegraphics[]{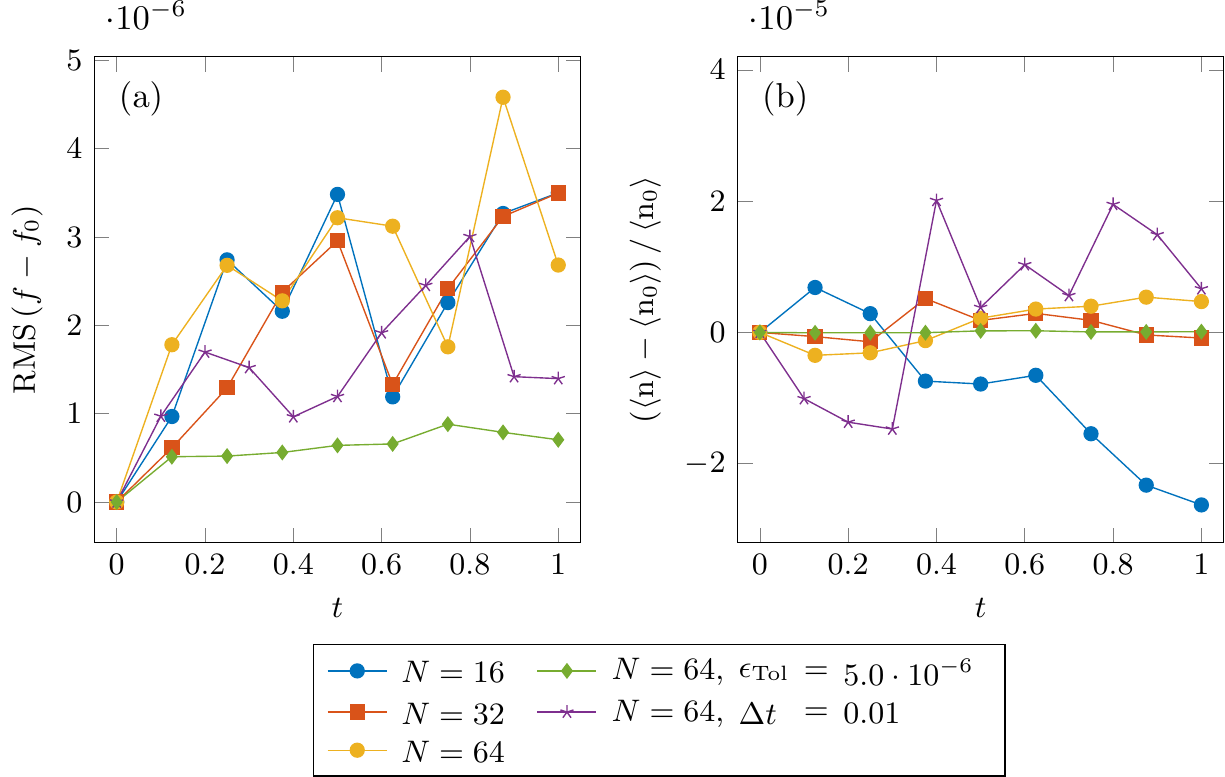}
\caption{(a) Temporal variability of the RMSE between the computed PDF $f$ and the initial PDF $f_{0}$. 
Unless otherwise mentioned in the legend $\Delta t = 0.025$ and $\epsilon_{\txt{Tol}} = 5.0 \cdot 10^{-5}$. 
The solution error is relatively independent from the
number of collocation points per dimension, $N$. However, reducing the time step $\Delta t$ and, to
a larger extent, the convergence criterion $\epsilon_{\txt{Tol}}$ significantly reduces the solution error. (b) Mass conservation as function of time $t$. 
The results show a clear improvement in mass conservation going from
$N = 16$ to $N = 32$, but not from $N = 32$ to $N = 64$.
Decreasing the time step to $\Delta t = 0.01$ does not reduce mass loss, while
reducing the convergence criterion to $\epsilon_{\txt{Tol}} = 5.0 \cdot 10^{-6}$ significantly
improves mass conservation.}
\label{fig:errorConst}
\end{figure}

Another metric of the accuracy of our steady-state
equilibrium solution, the root-mean-square error (RMSE) of
the solution is shown in figure~\ref{fig:errorConst} as function of time. The RMSE is defined as 
\begin{equation}
  \txt{RMS}(f - f_0)
= 
  \sqrt{ \frac{ 1 }{ N^{6}} \sum_{i=1}^{N^{6}} (f_{i} - f_{0,i})^2},
\end{equation}
where $f_{0}$ is the initial condition. The simulation was performed from $N
= 16$ to $N = 64$ collocation points per dimension. However, the number of collocation points does not
have a significant effect on the accuracy of the algorithm (figure~\ref{fig:errorConst}a). Decrease in the
time step, from $\Delta t = 0.025$ to  $\Delta t = 0.01$, does not have much
effect either. On the other hand, reducing the convergence criterion for $\beta$
from $\epsilon_{\txt{Tol}} = 5.0 \cdot 10^{-5}$ to  $\epsilon_{\txt{Tol}} = 5.0
\cdot 10^{-6}$ significantly lowers the error. This suggests the presence of a
bottleneck in increasing the accuracy of the simulation caused by the
convergence criterion. We show in section~\ref{sec:transient} that the
bottleneck for higher accuracy can depend on the physical system.

Since the continuity equation is used as an additional constraint, we explore 
the behavior of the RMSE of the spatially averaged density $\avr{n}$.
Figure~\ref{fig:errorConst}b shows that our algorithm satisfies mass conservation even at the lowest number of collocation points $N = 16$. As the number of collocation points increases from $N = 16$ to $N = 32$,  the
error in the mass conservation is significantly reduced. However, the further
increase in the number of collocation points, from $N = 32$ to $N = 64$, hardly
has an effect. Also, decreasing the time step from $\Delta t = 0.025$ to
$\Delta t = 0.01$ does not reduce the error. However, as was also observed in
figure~\ref{fig:errorConst}a, reducing the convergence criterion for $\beta$
from $\epsilon_{\txt{Tol}} = 5.0 \cdot 10^{-5}$ to  $\epsilon_{\txt{Tol}} = 5.0
\cdot 10^{-6}$ significantly reduces the amount of mass loss. This again
confirms that the convergence criterion serves as a bottleneck in increasing the simulation accuracy. This finding highlights the importance of picking a convergence-criterion value that satisfies the
desired balance between available computational resources and accuracy.
A smaller convergence criterion reduces the error, but results in
more iterations to reach convergence and, potentially, in a higher rank of the
solution, resulting in higher memory usage.

\begin{figure}[htb]
\center
\includegraphics[]{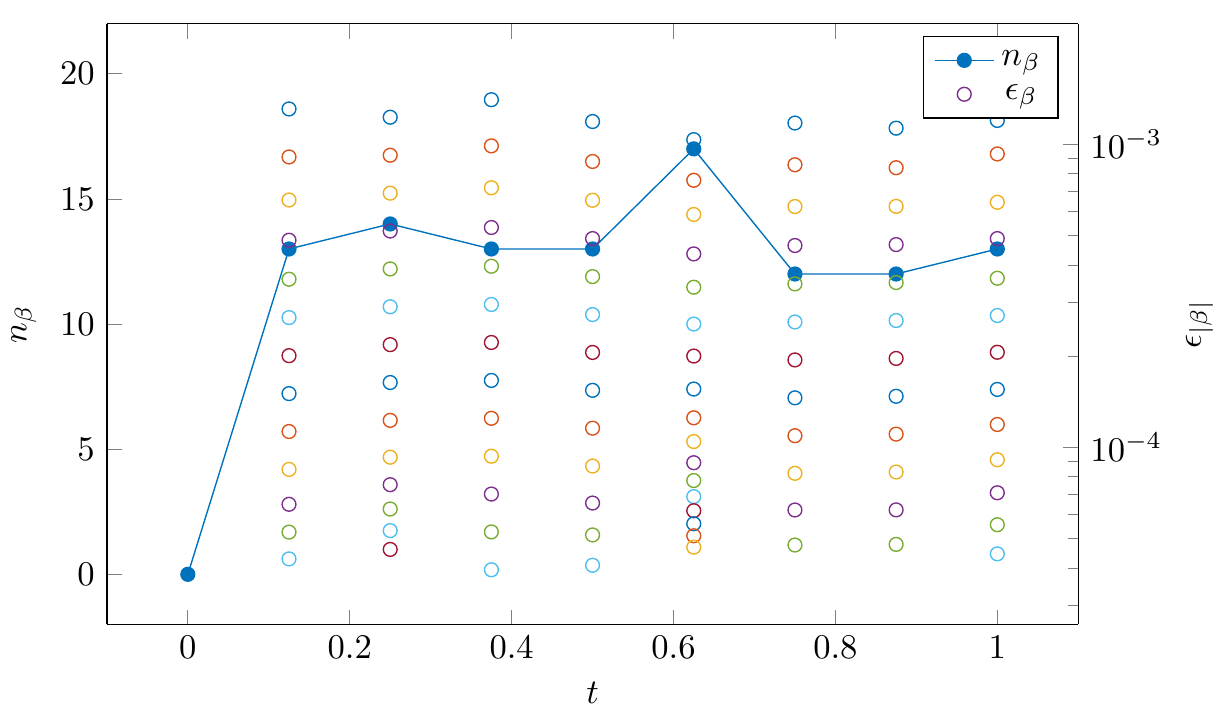}
\caption{Temporal variability of the umber of iterations, $n_{\beta}$, and convergence, $\epsilon_{|\beta|}$. The number of collocation points per dimension is $N=64$. SInce the same amplitude of random noise to $\beta$ is injected at every time step, the number of iterations is nearly constant in time. The
right axis has logarithmic scale, and the distance between each iteration
decreases as convergence is reached. This suggest that convergence follows an
exponential decay.}
\label{fig:betaConst}
\end{figure}

The number of iterations, $n_{\beta}$, needed to reach convergence throughout the simulation is reported in figure~\ref{fig:betaConst} (the left vertical axis) as function of time $t$. Because the same amount of random noise is added to $\beta$ before starting the ALS algorithm at every time step, the number of iterations is nearly
constant. Adding a smaller amount of random noise might reduce the number of iterations, but our numerical experiments revealed that doing so causes the ALS algorithm to get stuck in a local minimum and precludes 
the residual of the continuity equation, $\epsilon_{|\beta|}$, from being properly minimized. Figure~\ref{fig:betaConst} (the right vertical axis) shows that, as the convergence is reached,
the difference in the residual between the successive iterations gets smaller. This
suggest an exponential decay towards the minimum residual that can be
reached before the rank of the solution needs to be increased.

\begin{figure}[htbp]
\center
\includegraphics[]{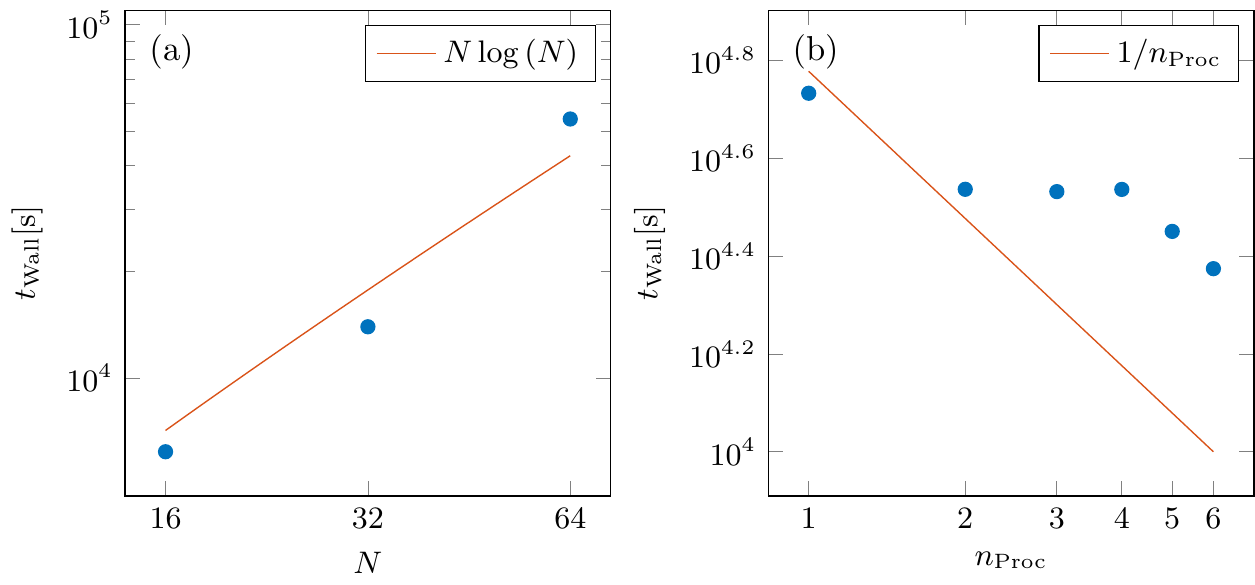}
\caption{ Scaling of the wall time $t_{\txt{Wall}}$ with (a) the
number of collocation points per dimension, $N$, and (b) the number of processors, $n_{\txt{Proc}}$. The
calculations to determine the scaling of the wall time as function of the number
of collocation points were performed on a single core.}
\label{fig:performConst}
\end{figure}

The computational efficiency of our code is reported in figure~\ref{fig:performConst}. The left frame
 shows the scaling of the wall time, $t_{\txt{Wall}}$, with the
number of collocation points per dimension, $N$, on one CPU core. 
The performance of the code is close to $N \log{ \brc{N}}$ in the range of the
explored collocation points. The most time is spent in the LSQR \citep{paige1982} subroutine, which is used to implicitly solve
for $\beta$ for each dimension during every iteration of the ALS procedure.
This suggest that the code could be further optimized by replacing the existing LSQR algorithm with its more efficient implementation, e.g.,~\citep{huang2013}. The right frame of figure~\ref{fig:performConst} exhibits the scaling of the wall time, $t_{\txt{Wall}}$, with the number of processors, $n_{\txt{Proc}}$. The curve $1/n_{\txt{Proc}}$ represents the ideal scaling without any communication overhead. Even though the different dimensions can all be solved for independently, the scaling of our code with the number of processors is quite poor. This
suggests that the code can be optimized further by minimizing the
communication between different CPU cores. One way to approach this would be to
rewrite the code around a specialized parallel processing MPI library and have
finer control over both data communication between processor cores and protocol selection.

\FloatBarrier

\subsection{Relaxation to statistical equilibrium in three spatial dimensions} 
\label{sec:transient}

\begin{figure}
\center
\includegraphics[width=0.8\textwidth]{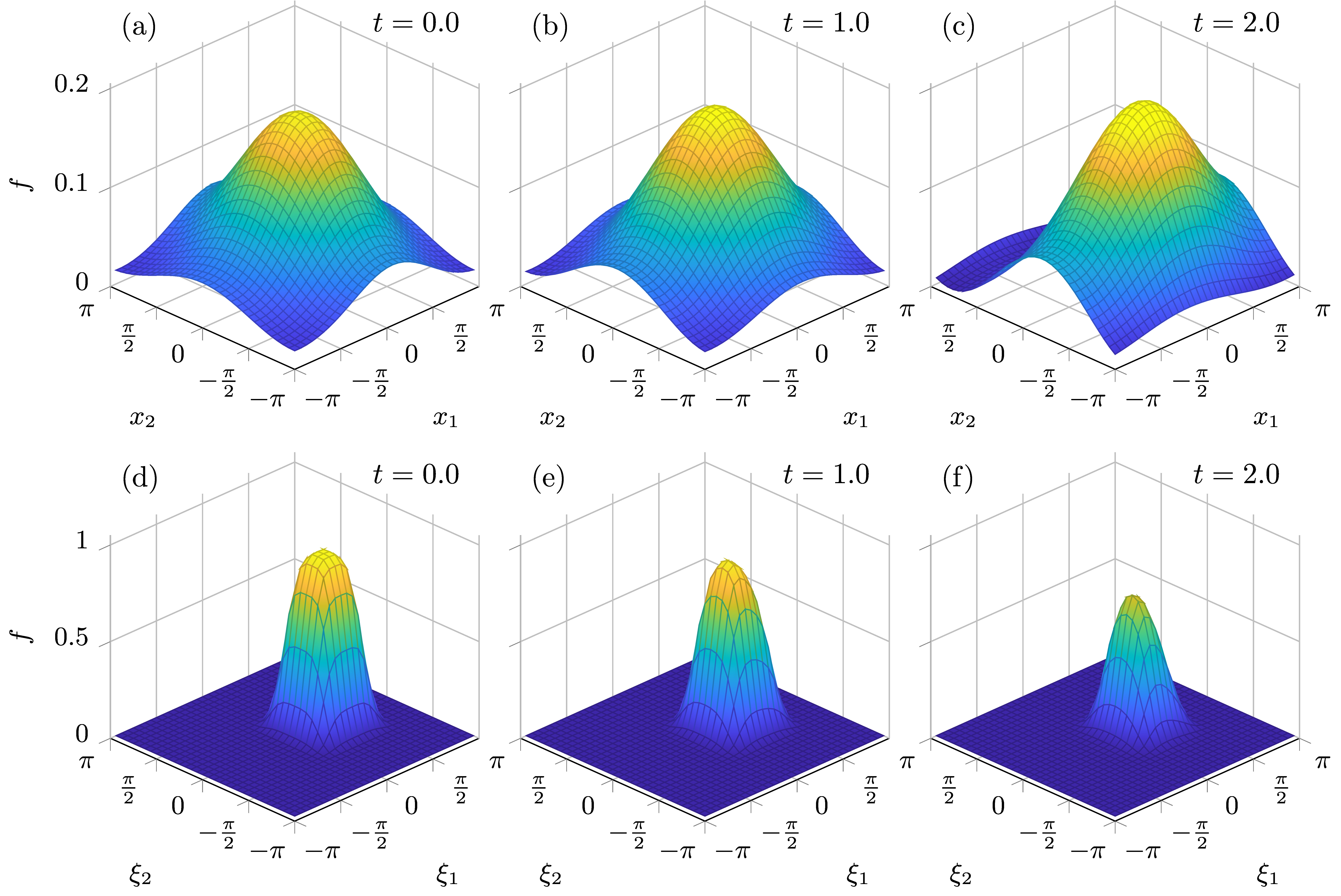}
\caption{Temporal evolution of the PDF $f(\mathbf x, \bm \xi, t)$ in hyper-planes $(\mathbf x; \bm \xi) = 
(x_1,x_2,0; \mathbf 0)$ (top row)  and $(\mathbf 0; \xi_1,\xi_2,0)$ (bottom row). The number of
collocation points per dimension is $N=32$. The color-coding is
consistent among the different frames in a row. 
The PDF $f$ evolves from its initial, far-from-equilibrium state~\eqref{eqn:initialC} towards its equilibrium state~\eqref{eqn:maxwellBoltzmann}.
}
\label{fig:fuv}
\end{figure}

\begin{figure}[htbp]
\center
\includegraphics[width=0.8\textwidth]{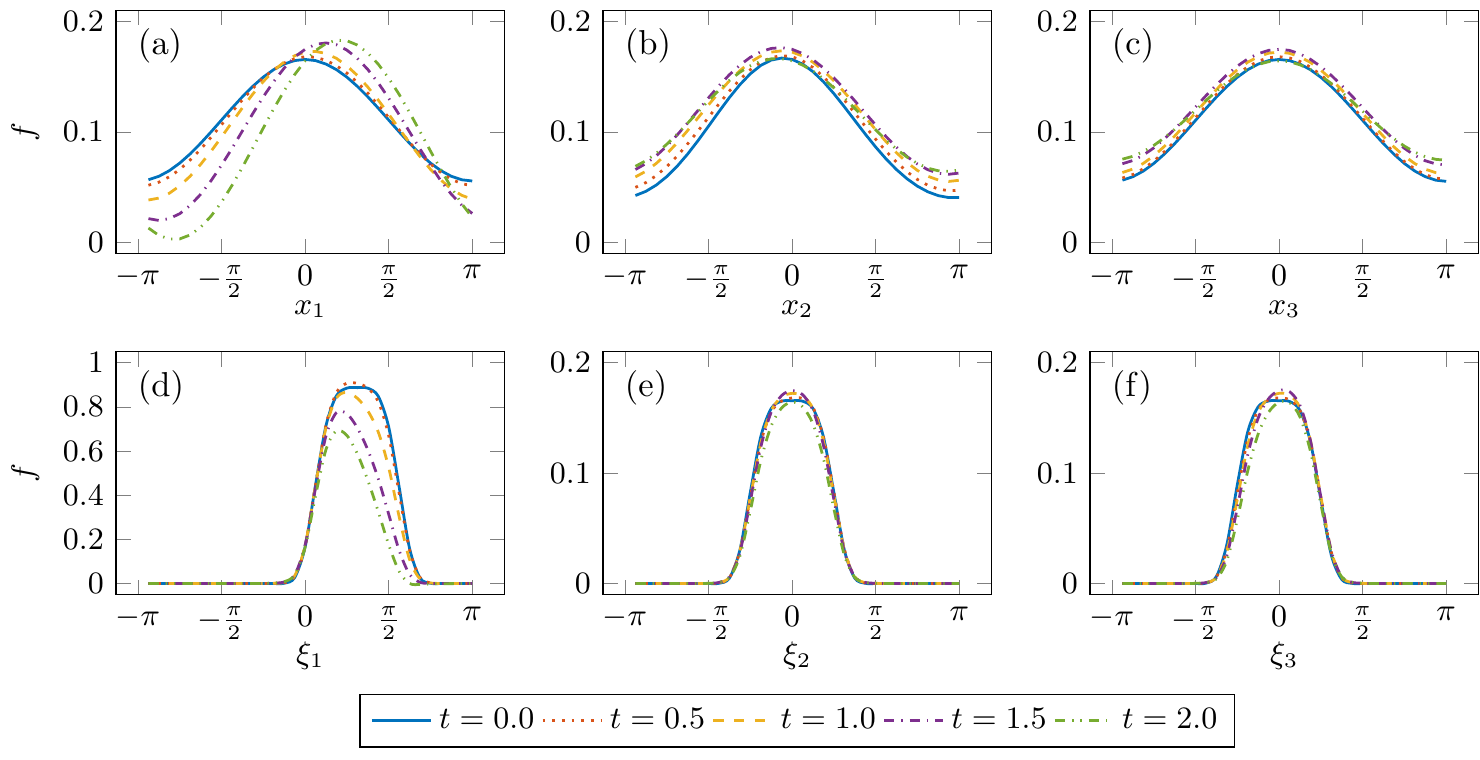}
\caption{ Temporal evolution of the PDF $f(\mathbf x, \bm \xi, t)$ in hyper-planes 
(a) $(\mathbf x; \bm \xi) = (x_1,0,0; \mathbf 0)$, 
(b) $(\mathbf x; \bm \xi) = (0,x_2,0; \mathbf 0)$, 
(c) $(\mathbf x; \bm \xi) = (0,0,x_3; \mathbf 0)$, 
(d) $(\mathbf x; \bm \xi) = (\mathbf 0;\xi_1,0,0)$, 
(e) $(\mathbf x; \bm \xi) = (\mathbf 0;0,\xi_2,0)$, and
(f) $(\mathbf x; \bm \xi) = (\mathbf 0;0,0,\xi_3)$. 
The number of collocation points per dimension is $N=32$. 
The PDF $f$ evolves from its initial, far-from-equilibrium state~\eqref{eqn:initialC} towards its equilibrium state~\eqref{eqn:maxwellBoltzmann}.
}
\label{fig:pdfDecay}
\end{figure}

In this section, we study relaxation to statistical 
equilibrium predicted by the dimensionless Boltzmann-BGK model
\eqref{BOBGK_dimensionless}--\eqref{eqn:temperatureDim}, subject to the initial condition 
\begin{subequations}\label{eqn:initialC}
\begin{equation}
f (\mathbf x,\bm \xi,0) = W f_{1} (x_{1},\xi_{1},0) f_{2} (x_{2},\xi_{2},0) f_{3} (x_{3},\xi_{3},0)
\end{equation}
with
\begin{align}
f_i (x_i,\xi_i,0) &= \frac{\sqrt[3]{n_{0}}}{\sqrt{2 \pi T_{0}/\txt{Bo}}} \exp\left[-\frac{\txt{Bo}}{2 T_{0}} (U_{i,0}-\xi_i)^4\right], \qquad i=1,2,3
\end{align}
\end{subequations}
and $n_0 = \prod_{i=1}^3 (0.5 \cos x_i +1)$, $T_0 = 0.0025 \cos(x_1)$, $U_{1,0} = 1 + 0.025 \sin(x_2-1)$, $U_{2,0} = 0$, and $U_{3,0} = 0.025 \sin(x_1-2)$. The integral over the PDF is
normalized to 1 by computing the value of $W$. The difference between the initial PDF and the local
equilibrium PDF causes the Boltzmann equation to evolve while the
fluctuations in the initial fields are introduced to show the code's ability to operate
away from global equilibrium. In this experiment, the Knudsen number is set to $\txt{Kn} = 10$. Figure~\ref{fig:fuv} shows the PDF $f(\mathbf x, \bm \xi, t)$ in the ($x_{1}$-$x_{2}$) and ($\xi_{1}$-$\xi_{2}$) hyper-planes. The PDF $f$ evolves from its initial state far from equilibrium to the equilibrium Maxwell-Boltzmann PDF $f_\text{eq}$. Figure~\ref{fig:pdfDecay} further elucidates this dynamics by exhibiting temporal snapshots of the PDF $f(\mathbf x, \bm \xi, t)$ in hyper-planes $(\mathbf x; \bm \xi) = (x_1,0,0; \mathbf 0)$, $(0,x_2,0; \mathbf 0)$, $(0,0,x_3; \mathbf 0)$, $(\mathbf 0;\xi_1,0,0)$, $(\mathbf 0;0,\xi_2,0)$, and $(\mathbf 0;0,0,\xi_3)$. 

\begin{figure}[htbp]
\center
\includegraphics[width=0.8\textwidth]{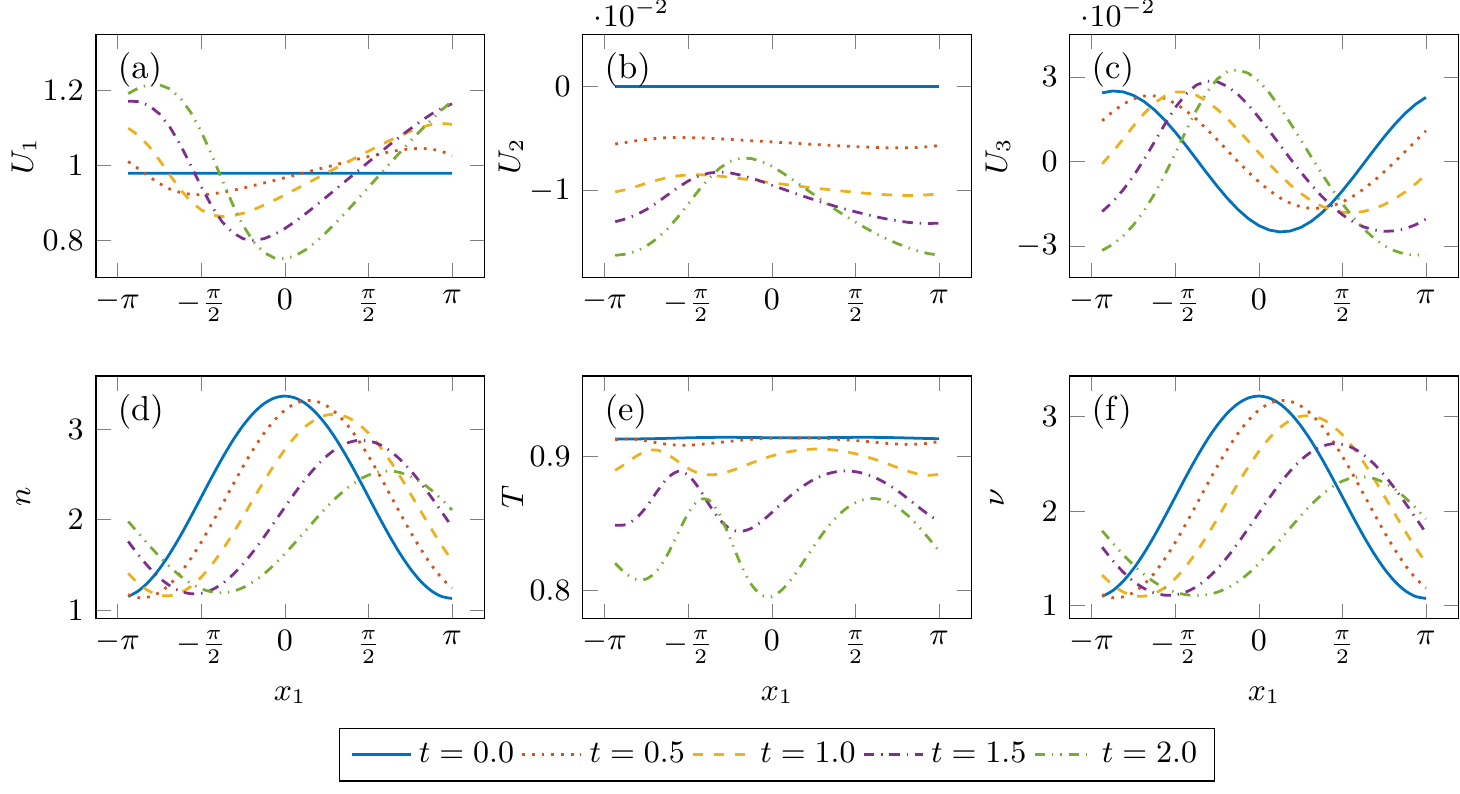}
\caption{Temporal evolution of (a--c) the velocity components $U_i\mathbf x,t)$, (d) number density $n(\mathbf x,t)$, (e) temperature $T(\mathbf x,t)$, and (f) the collision frequency $\nu(\mathbf x,t)$, all evaluated at the hyper-plane $\mathbf x = (x_1,0,0)$. The number of collocation points per dimension is $N=32$. 
}
\label{fig:fieldsDecay}
\end{figure}

Finally, figure~\ref{fig:fieldsDecay} exhibits temporal evolution of the number density $n(\mathbf x,t)$, velocity $\mathbf U(\mathbf x,t)$, temperature $T(\mathbf x,t)$, and the collision frequency $\nu(\mathbf x,t)$, all evaluated at the hyper-plane $\mathbf x = (x_1,0,0)$. As expected, the magnitude of the velocity components $U_{2}$ and $U_{3}$ is about $100$ times smaller than that of the $U_{1}$ component (frames a--c). The effect of the flow in the $x_{1}$ direction is seen in frame (d), where the density profile shifts to the right as time
progresses; in addition, the height of the density profile decreases as the mass is redistributed by diffusion. While not discernible from these figures, the code does suffer from a small amount of mass loss of about $2\%$ per
unit time. This is most likely due to a combination of small truncation errors,
convergence tolerances, and time step size. Frame (e) shows evolution of the temperature $T(x_1,\cdot,t)$ from its initial nearly uniform state. Increasing velocity fluctuations drive the increase in temperature fluctuations. The
collision frequency (frame f) is computed directly from the density and
temperature fields according to \ref{collfreq}. Overall all, the PDF and its moments show the expected physical behavior and evolve towards the equilibrium Maxwell-Boltzmann distribution.

\section{Discussion and conclusions}
\label{sec:concl}

We demonstrated the ability of canonical polyadic tensor decomposition 
to solve the BGK approximation of the six-dimensional Boltzmann transport equation with variable density, velocity, temperature, and collision frequency fields. This extends the usage of this method from only
fully separable differential operators, to partially separable operators. The dependent variables in the Boltzmann equation are computed via the pseudo-spectral method with collocation points. The different dimensions are
solved using a parallel ALS algorithm. Our numerical experiments show
that the code is capable of simulating both a system's steady state and its temporal evolution starting from a state far away from equilibrium. The highest simulation accuracy is achieved by identifying a correct
bottleneck: in the steady-state simulations the limiting factor is the convergence
tolerance. The code's performance scales as $N \log(N)$ with the number of collocation points per dimension, $N$.

In future work, we will further optimize the code, implement different boundary conditions, and explore the feasibility of using the code with other collision operators. Significant speedup can be achieved by rank reduction of the  collision operator. The currently used reconstruction of the collision operator via multiplication
of its different components can result in a high-rank operator, which can become
degenerate. Reducing the rank of this tensor, while maintaining accuracy, is a
topic that needs further study. Another aspect where the code can be improved is
parallelization, especially when using the CNLF-ALS method to solve problems of
 dimensionality higher than that of the Boltzmann-BGK equation. One way to approach this
would be to rewrite the code around a specialized parallel processing MPI
library and have finer control over data communication between processor
cores and protocol selection.
 
Generalizations to other types of boundary conditions would require an in-depth investigation of the trial function behavior. The choice of a right trial function for a certain system is typically determined by one of the two 
approaches. First, one selects a trial function that obeys the desired
boundary conditions and then sums over the trial functions to reconstruct the
solution to a PDE. Second, one selects a trial function that captures the PDE solution and then adds trial functions to reconstruct the boundary condition. Figure~\ref{fig:pdfDecay} shows that the six-dimensional Boltzmann  equation subject to periodic boundary conditions can be efficiently solved by using the discrete Fourier series. The presence of, e.g., a wall would introduce the Maxwell boundary condition~\cite{cercignani1988,maxwell1878}. The latter consists of two parts: one represents the reflection of particles from the wall and the other
represents absorption and emission of particles on the wall. Since there is no
known trial function to either solve the Boltzmann equation in the bulk or the
Maxwell boundary condition on the wall, another method of the trial-function selection is needed. Such methods include the tau method~\citep{lanczos1956}, the penalty
approach~\citep{hesthaven2000}, and the mixed method~\citep{shuleshko1959,snyder1964,zinn1968}. 
They have been used to implement boundary condition for
the weighted residuals and spectral methods~\citep{gottlieb1977,hesthaven2000},
but have not yet been applied in the tensor decomposition setting.

While the boundary conditions are very
important for engineering applications, some fundamental questions remain to be resolved as well. The empirical evidence shows convergence the tensor algorithms, but its theoretical proof is lacking.
That is largely due to two factors. First, it is generally assumed that increasing the rank of the solution improves its accuracy and in, some cases, tensor decomposition was shown to be exponentially
more efficient than one would expect a priori~\citep{beylkin2009}. However, there
is no mathematical proof which indicates how high the rank needs to be to reach
a certain accuracy and how fast the solution converges as a function of its
rank. This question is very important for the speed of  tensor decomposition, because the higher the rank of the operators and the solution of the
Boltzmann equation, the slower the computation becomes.
Second, from one time step to the next convergence is not
guaranteed. The ALS method implemented in our code is a variation on the
Gauss-Seidel method whose convergence can be proven for some specific cases
\citep{uschmajew2012}. However, only the Jacobi method can decouple the different dimensions and parallelize the algorithm. How the adoption of this method would change convergence remains to be investigated.

\vspace{0.5cm}
\noindent 
{\bf Acknowledgements} 
This research was supported in part by the U.S. Army Research 
Office (ARO) grant W911NF1810309 awarded to 
Daniele Venturi, and by the Air Force Office of Scientific 
Research (AFOSR) grant FA9550-18-1-0474 awarded to 
Daniel Tartakovsky.

\bibliographystyle{elsart-num}

\pagebreak

\begin{center}
  {\Large Supplementary Materials to ``Tensor methods for the Boltzmann-BGK equation''}\\[.5cm]
  Arnout M. P. Boelens$^{\txt{a}}$, Daniele Venturi$^{\txt{b}}$ Daniel M. Tartakowski$^{\txt{a}}$\\[.2cm]
  {\itshape \small $^{\txt{a}}$ Department of Energy Resources Engineering, Stanford University, Stanford, CA 94305 \\
  $^{\txt{b}}$ Department of Applied Mathematics, UC Santa Cruz, Santa Cruz, CA 95064 \\[1.0cm]}
\end{center}

\hrulefill

\hrulefill

\setcounter{equation}{0}
\setcounter{figure}{0}
\setcounter{table}{0}
\setcounter{page}{1}
\makeatletter
\renewcommand{\theequation}{S\arabic{equation}}
\renewcommand{\thefigure}{S\arabic{figure}}
\renewcommand{\bibnumfmt}[1]{[S#1]}
\renewcommand{\citenumfont}[1]{S#1}

Equations~(15)--(19) are solved with a Matlab~\citep{mathworks2017} code that uses the Matlab Tensor Toolbox~\citep{bader2017}.
One of the core elements of the code is a special case of the Tucker operator,
the so-called ``Kruskal operator''~\citep{kolda2006,kruskal1977}. In three
dimensions this operator is defined as
\begin{equation}
  a_{i,j,k} 
= 
  \sum_{r_{a}=1}^{R_{a}} \lambda_{r_{a}} \; a_{1,i,r_{a}} \; a_{2,j,r_{a}} \; a_{3,k,r_{a}} \txt{ for all $i,j,k$},
\end{equation}
where $R_{a}$ is the rank of the operator. In general, $\lambda_{r}$ is used as a
prefactor to normalize the matrices $a_{1,i,r}, \dots ,a_{6,n,r}$, but in our
code $\lambda_{r} = 1$ for simplicity. The code uses a number of basic
operations: the \textproc{mtimesk} subroutine of Algorithm~\ref{alg:g} performs simple multiplication of
two Kruskal operators $a_{i,j,k}$ and $b_{i,j,k}$. For example, for all $i,j,k$,
\begin{align}
  \textproc{mtimesk} \brc{ a_{i,j,k},b_{i,j,k}} =
 \sum_{r_{a}=1}^{R_{a}} \sum_{r_{b}=1}^{R_{b}} \lambda_{r_{a}} \lambda_{r_{b}} \; a_{1,i,r_{a}} b_{1,i,r_{b}} \; a_{2,j,r_{a}} b_{2,j,r_{b}} \; a_{3,k,r_{a}} b_{3,k,r_{b}},
\end{align}
creating a new Kruskal operator of rank $R=R_{a} R_{b}$. The
\textproc{mtimeskBlock} subroutine also performs multiplication but does so in
blocks. In the case of $2 \times 2$ blocks of length $R_{a}'$ and
$R_{b}'$, 
\begin{multline}
\textproc{mtimeskBlock} \brc{a_{i,j,k},b_{i,j,k},R_{a},R_{b},R_{a}',R_{b}'} = \\ 
\begin{aligned}
 & \sum_{r_{a}=1}^{R_{a}'}       \sum_{r_{b}=1}^{R_{b}'}       \lambda_{r_{a}} \lambda_{r_{b}} \; a_{1,i,r_{a}} b_{1,i,r_{b}} \; a_{2,j,r_{a}} b_{2,j,r_{b}} \; a_{3,k,r_{a}} b_{3,k,r_{b}} \\
+& \sum_{r_{a}=1}^{R_{a}'}       \sum_{r_{b}=R_{b}'+1}^{R_{b}} \lambda_{r_{a}} \lambda_{r_{b}} \; a_{1,i,r_{a}} b_{1,i,r_{b}} \; a_{2,j,r_{a}} b_{2,j,r_{b}} \; a_{3,k,r_{a}} b_{3,k,r_{b}} \\
+& \sum_{r_{a}=R_{a}'+1}^{R_{a}} \sum_{r_{b}=1}^{R_{b}'}       \lambda_{r_{a}} \lambda_{r_{b}} \; a_{1,i,r_{a}} b_{1,i,r_{b}} \; a_{2,j,r_{a}} b_{2,j,r_{b}} \; a_{3,k,r_{a}} b_{3,k,r_{b}} \\
+& \sum_{r_{a}=R_{a}'+1}^{R_{a}} \sum_{r_{b}=R_{b}'+1}^{R_{b}} \lambda_{r_{a}} \lambda_{r_{b}} \; a_{1,i,r_{a}} b_{1,i,r_{b}} \; a_{2,j,r_{a}} b_{2,j,r_{b}} \; a_{3,k,r_{a}} b_{3,k,r_{b}} \txt{ for all $i,j,k$}.
\end{aligned}
\end{multline}
The subroutines \textproc{krepmat} and \textproc{krepelem} repeat the whole
Kruskal object or its individual elements, respectively. For instance, for all $i,j,k$,
\begin{align}
  \textproc{krepmat} \brc{a_{i,j,k},2} 
=
   \sum_{r_{a}=1}^{R_{a}} \lambda_{r_{a}} \; a_{1,i,r_{a}} \; a_{2,j,r_{a}} \; a_{3,k,r_{a}} 
+ \sum_{r_{a}=1}^{R_{a}} \lambda_{r_{a}} \; a_{1,i,r_{a}} \; a_{2,j,r_{a}} \; a_{3,k,r_{a}} 
\end{align}
and
\begin{align}
  \textproc{krepelem} \brc{a_{i,j,k},2} = 
 \sum_{r_{a}=1}^{R_{a}} \lambda_{r_{a}} \; a_{1,i,r_{a}} \; a_{2,j,r_{a}} \; a_{3,k,r_{a}} + \lambda_{r_{a}} \; a_{1,i,r_{a}} \; a_{2,j,r_{a}} \; a_{3,k,r_{a}}.
\end{align}
The \textproc{kreshape} subroutine extends a Kruskal operator to a higher dimension. To add the fourth dimension to $a_{i,j,k}$, one writes
\begin{equation}
  \textproc{kreshape} \brc{a_{i,j,k},[1 \; 1 \; 0 \; 1]}
=
  \sum_{r_{a}=1}^{R_{a}} \lambda_{r_{a}} \; a_{1,i,r_{a}} \; a_{2,j,r_{a}} \; b_{1,l,r_{a}} \; a_{3,k,r_{a}}, \quad \txt{ for all $i,j,k,l$}.
\end{equation}
In this example, $a_{3,k,r_{a}}$ becomes the fourth dimension and the third
dimension equals $b_{1,l,r_{a}} = 1$ for all values of $l$ and $r_{a}$.

\begin{algorithm}[htbp]
\caption{computeG subroutine}
\label{alg:g}
\begin{algorithmic}[1]
\Function{computeG}{$A,\beta,R,R_{\txt{Exp}}$}
\State
\State{$\beta \gets \Call{krepelem}{\beta,R_{\txt{Exp}}}$} \Comment{Repeat elements of beta}
\State
\State{$\beta \gets \Call{computeF}{\beta,[0 \; 0 \; 0 \; 1 \; 1 \; 1]}$} \Comment{iFFT where A works in real space}
\State
\State{$\beta \gets \Call{mtimeskBlock}{A,\beta,R_{\txt{A}},R,1,R_{\txt{Exp}}}$} \Comment{Apply operator A}
\State
\State{$G \gets \Call{computeF}{\beta,[1 \; 1 \; 1 \; 0 \; 0 \; 0]}$} \Comment{iFFT where A works in Fourier space}
\State 
\State \Return{$G$}
\EndFunction
\end{algorithmic}
\end{algorithm}

Algorithm~\ref{alg:als} shows the Parallel Alternating Least Squares (ALS)
algorithm used to solve the Boltzmann transport equation. The subroutine
\textproc{Initialization} initializes all variables to run the code. This
includes variables like the ones listed in table~1, but also the operators
$A$, $A^{+}$, $A^{-}$, $\txt{Exp}$, $\txt{Exp}_{A}$, and $\txt{Exp}_{{A}^{+}}$.
The operators $A$, $A^{+}$, and $A^{-}$ are the respective Kruskal operator
representations of the operators introduced in section~3.3. For
these operators, multiplication with $\xi_{k}$ is represented in real space and
derivatives are represented in Fourier space to achieve spectral accuracy. This
means that to apply these operators to a function, the velocity space, i.e., the
first three dimensions, of this function have to be in physical space and the
velocity space, i.e., the last three dimensions, have to be in Fourier space.
This can be seen in the \textproc{computeG} subroutine in Algorithm~\ref{alg:g}
that is used throughout the code.

\begin{algorithm}[htbp]
\caption{Parallel Alternating Least Squares (ALS) algorithm}
\label{alg:als}
\begin{algorithmic}[1]
\Procedure{Main}{}
\State{\Call{Initialization}{}} \Comment{Load variables and allocate matrices}
\State{$\beta_{\txt{New}} = \beta_{\txt{Now},0}$} \Comment{Set initial condition}
\State{$\beta_{\txt{Now}} = \beta_{\txt{Now},0}$}
\State{$\beta_{\txt{Old}} = \beta_{\txt{Old},0}$}
\State
\For{$n \gets 1:n_{\txt{max}}$}
\State{$C \gets \Call{computeArrayC}{\beta_{\txt{Now}}}$} \Comment{Compute collision operator}
\State
\State{$\beta_{\txt{New}} \gets \Call{randBeta}{\beta_{\txt{New}}}$} \Comment{Add some random noise}
\State{$\epsilon_{| \beta |} \gets \epsilon_{\txt{Tol}} + 10^{6}$} \Comment{Reset stop criterion}
\While{$\epsilon_{| \beta |} > \epsilon_{\txt{Tol}}$}
\State{$\beta_{\txt{Int}} \gets \beta_{\txt{New}}$} \Comment{Set intermediate value of beta}
\State
\State{$N_{0} \gets \Call{computeArrayN0}{A^{+},A^{-},\txt{Exp}_{A^{+}},\beta_{\txt{Old}},\beta_{\txt{New}}}$}
\State{$O_{0} \gets \Call{computeArrayO0}{C    ,A^{+},\txt{Exp}_{A^{+}},\beta_{\txt{New}}}$}
\State{$\gamma_{0} \gets N_{0} + 2 \; \Delta t \; O_{0}$}
\State
\State{$\gamma_{1} \gets \Call{computeArrayN1}{\txt{Exp},A,\txt{Exp}_{A},\beta_{\txt{Now}},\beta_{\txt{New}}}$}
\State
\ParFor{$d \gets 1:6$} \Comment{Iterate over dimensions}
\State{$M_{0} \brc{d} \gets \Call{computeArrayM0}{A^{+},\txt{Exp}_{A^{+}} \brc{d},\beta_{\txt{New}},d}$}
\State{$M_{1} \brc{d} \gets \Call{computeArrayM1}{\txt{Exp},\txt{Exp}_{A} \brc{d},A,\beta_{\txt{New}},d}$}
\EndParFor
\State
\ParFor{$d \gets 1:6$}
\State{$\beta_{\txt{New}} \brc{d} \gets \Call{computeBetaNew}{\beta_{\txt{New}} \brc{d}, M_{0} \brc{d}, \gamma_{0} \brc{d}, M_{1} \brc{d}, \gamma_{1} \brc{d}}$}
\EndParFor
\State
\State{$\beta_{\txt{New}} \gets \Call{kreal}{\beta_{\txt{New}}}$} \Comment{Only keep real part of solution}
\State{$\epsilon_{| \beta |} \gets \Call{computeNormBeta}{\beta_{\txt{Int}},\beta_{\txt{New}}}$} \Comment{Check for convergence}
\State
\If{$ \epsilon_{| \beta |} > \epsilon_{\txt{Tol}}$}
\State{$\beta_{\txt{New}} \gets \Call{computeBetaNewDelta}{\beta_{\txt{New}},\beta_{\txt{Int}}}$} \Comment{Update}
\Else
\State{$\beta_{\txt{New}},\beta_{\txt{Now}} \gets \Call{computeRAW}{\beta_{\txt{New}},\beta_{\txt{Now}},\beta_{\txt{Old}}}$} \Comment{Apply RAW filter}
\EndIf
\State
\EndWhile
\State{$\beta_{\txt{Old}} = \beta_{\txt{Now}}$} \Comment{Update for next time step}
\State{$\beta_{\txt{Now}} = \beta_{\txt{New}}$}
\EndFor
\EndProcedure
\end{algorithmic}
\end{algorithm}

For reasons that will become clear below the
first step in this code is to repeat the elements of $\beta R_{\txt{Exp}}$
times. After this, operator $A$ is applied. This involves first applying the
inverse Fourier transform to the velocity space, then applying operator $A$, and
finally applying the inverse Fourier transform to position space, completing the
procedure. $\txt{Exp}$ is a 2-D Kruskal operator representing the trial function
$\phi_{s}$, $\txt{Exp}_{A}$ is the Kruskal operator representing operator $A$
acting on trail function $\phi_{s}$, and $\txt{Exp}_{{A}^{+}}$ is the Kruskal
operator representing operator $A^{+}$ acting on trail function $\phi_{s}$. The
reason to make one Kruskal operator of operators $A$ and $A^{+}$ acting on the
trial function is that this reduces the loss of accuracy that comes with
multiplying Kruskal operators. \textproc{Initialization} also computes initial
guesses for variables which need to be calculated to compute the collision
operator $C$. These are the variables $UU_{0}$, $VV_{0}$, $WW_{0}$,
$\txt{exp}_{U,0}$, $\txt{exp}_{V,0}$, $\txt{exp}_{W,0}$, and $\nu_{0}$ which are
found in Algorithm~\ref{alg:c}. In addition to these variables
$\beta_{\txt{New}}$, $\beta_{\txt{Now}}$, and $\beta_{\txt{Old}}$ are initialized.

\begin{algorithm}[htbp]
\caption{computeArrayC subroutine}
\label{alg:c}
\begin{algorithmic}[1]
\Function{computeArrayC}{$\beta_{\txt{Now}}$}
\State{$f \gets \Call{computeF}{\beta_{\txt{Now}},[1 \; 1 \; 1 \; 1 \; 1 \; 1]}$} \Comment{Perform iFFT on all dimensions}
\State{$n_{\txt{Eq}} \gets \Call{computeDensity}{f}$} \Comment{Compute density field}
\State{$U,V,W \gets \Call{computeVelocities}{u, v, w, n_{\txt{Eq}}, f}$} \Comment{Compute average velocities}
\State
\ParFor{$n \gets 1:N^{4}$} \Comment{Iterate over collocation points in non-nested loop}
\State{$n_{1} \gets \Call{mod}{\repl{(((}{} n-1 \repl{)/N-(n_{2}-1))/N-(n_{3}-1)))/N}{},N} + 1$}
\State{$n_{2} \gets \Call{mod}{\repl{((}{} (n-n_{1})/N \repl{-(n_{2}-1))/N-(n_{3}-1)))/N}{},N} + 1$}
\State{$n_{3} \gets \Call{mod}{\repl{(}{} ((n-n_{1})/N-(n_{2}-1))/N \repl{-(n_{3}-1)))/N}{},N} + 1$}
\State{$n_{4} \gets \Call{mod}{(((n-n_{1})/N-(n_{2}-1))/N-(n_{3}-1))/N,N} + 1$}
\State
\State{$\repl{WW}{UU} \brc{n} \gets (\repl{W}{U} (n_{1},n_{2},n_{3})-\repl{w}{u} (n_{4}))^2$} \Comment{Create 4D blocks}
\State{$\repl{WW}{VV} \brc{n} \gets (\repl{W}{V} (n_{1},n_{2},n_{3})-\repl{w}{v} (n_{4}))^2$}
\State{$\repl{WW}{WW} \brc{n} \gets (\repl{W}{W} (n_{1},n_{2},n_{3})-\repl{w}{w} (n_{4}))^2$}
\EndParFor
\State{$\repl{WW}{UU} \gets \Call{reshape}{\repl{WW}{UU},[N,N,N,N]}$} \Comment{Reshape into $N \times N \times N \times N$ arrays}
\State{$\repl{WW}{VV} \gets \Call{reshape}{\repl{WW}{VV},[N,N,N,N]}$}
\State{$\repl{WW}{WW} \gets \Call{reshape}{\repl{WW}{WW},[N,N,N,N]}$}
\State
\State{$\repl{WW_{0}}{UU_{0}} \gets \Call{decompose}{\repl{WW}{UU},\txt{'init'},\repl{WW_{0}}{UU_{0}}}$} \Comment{Decompose into ktensors}
\State{$\repl{WW_{0}}{VV_{0}} \gets \Call{decompose}{\repl{WW}{VV},\txt{'init'},\repl{WW_{0}}{VV_{0}}}$}
\State{$\repl{WW_{0}}{WW_{0}} \gets \Call{decompose}{\repl{WW}{WW},\txt{'init'},\repl{WW_{0}}{WW_{0}}}$}
\State
\State{$T       \gets \Call{computeTemperature}{UU_{0}, VV_{0}, WW_{0}, n_{\txt{Eq}}, f}$} \Comment{Compute temperature}
\State{$\nu_{0} \gets \Call{computeFrequency}{n_{\txt{Eq}}, T, \nu_{0}}$} \Comment{Compute collision frequency}
\State
\State{$\repl{n_{\txt{Eq},1}}{T_{1}         } \gets 0.5/T$} \Comment{Compute some common factors}
\State{$\repl{n_{\txt{Eq},1}}{n_{\txt{Eq},1}} \gets n_{\txt{Eq}}^{1/3} \sqrt{T_{1}}/\sqrt{\pi/\txt{Bm}}$}
\State{$\repl{n_{\txt{Eq},1}}{T_{1}         } \gets T_{1} \txt{Bm}$}
\State
\ParFor{$n \gets 1:N^{4}$}
\State{$n_{1} \gets \Call{mod}{\repl{(((}{} n-1 \repl{)/N-(n_{2}-1))/N-(n_{3}-1)))/N}{},N} + 1$}
\State{$n_{2} \gets \Call{mod}{\repl{((}{} (n-n_{1})/N \repl{-(n_{2}-1))/N-(n_{3}-1)))/N}{},N} + 1$}
\State{$n_{3} \gets \Call{mod}{\repl{(}{} ((n-n_{1})/N-(n_{2}-1))/N \repl{-(n_{3}-1)))/N}{},N} + 1$}
\State{$n_{4} \gets \Call{mod}{(((n-n_{1})/N-(n_{2}-1))/N-(n_{3}-1))/N,N} + 1$}
\algstore{myalg}
\end{algorithmic}
\end{algorithm}

\begin{algorithm}[htbp]
\begin{algorithmic}[1]
\algrestore{myalg}
\State{$\repl{\txt{Exp}_{W}}{\txt{Exp}_{U}} \brc{n} \gets n_{\txt{Eq},1} \brc{n_{1},n_{2},n_{3}} \Call{Exp}{-T_{1} \brc{n_{1},n_{2},n_{3}} \brc{\repl{W}{U} (n_{1},n_{2},n_{3})-\repl{w}{u} (n_{4})}^2}$}
\State{$\repl{\txt{Exp}_{W}}{\txt{Exp}_{V}} \brc{n} \gets n_{\txt{Eq},1} \brc{n_{1},n_{2},n_{3}} \Call{Exp}{-T_{1} \brc{n_{1},n_{2},n_{3}} \brc{\repl{W}{V} (n_{1},n_{2},n_{3})-\repl{w}{v} (n_{4})}^2}$}
\State{$\repl{\txt{Exp}_{W}}{\txt{Exp}_{W}} \brc{n} \gets n_{\txt{Eq},1} \brc{n_{1},n_{2},n_{3}} \Call{Exp}{-T_{1} \brc{n_{1},n_{2},n_{3}} \brc{\repl{W}{W} (n_{1},n_{2},n_{3})-\repl{w}{w} (n_{4})}^2}$}
\EndParFor
\State
\State{$\repl{\txt{Exp}_{W}}{\txt{Exp}_{U}} \gets \Call{reshape}{\repl{\txt{Exp}_{W}}{\txt{Exp}_{U}}, [N,N,N,N]}$} \Comment{Reshape into $N \times N \times N \times N$ arrays}
\State{$\repl{\txt{Exp}_{W}}{\txt{Exp}_{V}} \gets \Call{reshape}{\repl{\txt{Exp}_{W}}{\txt{Exp}_{V}}, [N,N,N,N]}$}
\State{$\repl{\txt{Exp}_{W}}{\txt{Exp}_{W}} \gets \Call{reshape}{\repl{\txt{Exp}_{W}}{\txt{Exp}_{W}}, [N,N,N,N]}$}
\State
\State{$\repl{\txt{Exp}_{W,0}}{\txt{Exp}_{U,0}} \gets \Call{decompose}{\repl{\txt{Exp}_{W}}{\txt{Exp}_{U}}, \txt{`init'}, \repl{\txt{Exp}_{W,0}}{\txt{Exp}_{U,0}}}$} \Comment{Decompose into ktensors}
\State{$\repl{\txt{Exp}_{W,0}}{\txt{Exp}_{V,0}} \gets \Call{decompose}{\repl{\txt{Exp}_{W}}{\txt{Exp}_{V}}, \txt{`init'}, \repl{\txt{Exp}_{W,0}}{\txt{Exp}_{V,0}}}$}
\State{$\repl{\txt{Exp}_{W,0}}{\txt{Exp}_{W,0}} \gets \Call{decompose}{\repl{\txt{Exp}_{W}}{\txt{Exp}_{W}}, \txt{`init'}, \repl{\txt{Exp}_{W,0}}{\txt{Exp}_{W,0}}}$}
\State
\State{$\repl{\txt{Exp}_{W}}{\txt{Exp}_{U}} \gets \Call{kreshape}{\repl{\txt{Exp}_{W,0}}{\txt{Exp}_{U,0}}, [1 \; 1 \; 1 \; 1 \; 0 \; 0]}$} \Comment{Map 4-D ktensors to 6D ktensors}
\State{$\repl{\txt{Exp}_{W}}{\txt{Exp}_{V}} \gets \Call{kreshape}{\repl{\txt{Exp}_{W,0}}{\txt{Exp}_{V,0}}, [1 \; 1 \; 1 \; 0 \; 1 \; 0]}$}
\State{$\repl{\txt{Exp}_{W}}{\txt{Exp}_{W}} \gets \Call{kreshape}{\repl{\txt{Exp}_{W,0}}{\txt{Exp}_{W,0}}, [1 \; 1 \; 1 \; 0 \; 0 \; 1]}$}
\State
\State{$C \gets \Call{mtimesk}{\txt{Exp}_{U},\txt{Exp}_{V}}$} \Comment{Construct 6D ktensor}
\State{$C \gets \Call{mtimesk}{C,\txt{Exp}_{W}}$}
\State{$C \gets C - f$} \Comment{Difference between equilibrium and current distribution}
\State
\State{$\nu \gets \Call{kreshape}{\nu_{0},[1 \; 1 \; 1 \; 0 \; 0 \; 0]}$} \Comment{Map 3D ktensor to 6D ktensor}
\State
\State{$C \gets \Call{mtimesk}{\nu,C} \cdot 1/\txt{Kn}$}
\State
\State \Return{$C$}
\EndFunction
\end{algorithmic}
\end{algorithm}


The first subroutine executed for every time step is \textproc{computeArrayC}.
This subroutine is shown in more detail in Algorithm~\ref{alg:c}. In addition,
before starting the calculation of $\beta_{\txt{New}}$ some random noise is
added in subroutine \textproc{randBeta} and the convergence criterion
$\epsilon_{| \beta |}$ is reset. The addition of random noise assures that
the algorithm does not get stuck in a local minimum. Once inside the loop to
find the new value of $\beta_{\txt{New}}$ its current value is stored as
$\beta_{\txt{Int}}$. Various subroutines compute the matrices, $M_{0}$,
$\gamma_{0}$, $M_{1}$, and $\gamma_{1}$, needed to find the new value of
$\beta_{\txt{New}}$ using the subroutine \textproc{computeBetaNew}. This
subroutine uses the LSQR algorithm \citep{barrett1994} to find a solution for
$\beta_{\txt{New}}$ that simultaneously minimizes the residue for the Boltzmann
transport equation and the continuity equation. During every iteration only the
real part of the solution is kept. The subroutine \textproc{kreal} takes the
inverse Fourier transform of $\beta_{\txt{New}}$, discards the imaginary part of
the solution, and converts the solution back to Fourier space. The
convergence criterion is computed in the subroutine \textproc{computeNormBeta}.
Computation of the full residual at 
every iteration is computationally expensive; instead, the convergence is determined as
\begin{equation}
  \epsilon_{| \beta |}
=
  \max{\brc{  \frac{\nor{\beta_{\txt{New}} \brc{d} - \beta_{\txt{Int}} \brc{d}}}{\nor{\beta_{\txt{Int}} \brc{d}}}}}.
\end{equation}
The change in $\beta_{\txt{New}}$ is computed for every dimension $d$ and the
maximum value is chosen. When $\epsilon_{| \beta |} < \epsilon_{\txt{Tol}}$, the
while loop in Algorithm~\ref{alg:als} is allowed to exit. If the solution has not converged 
\textproc{computeBetaNewDelta} is used to calculate the new value of
$\beta_{\txt{New}}$ for the next iteration using 
\begin{equation}
  \beta_{\txt{New}} 
\gets
  \beta_{\txt{Int}} + \frac{{\beta_{\txt{New}} - \beta_{\txt{Int}}}}{\Delta
  \beta}
\end{equation}
with $\Delta \beta = 4$. It is important to gradually approach the location where the residual reaches a
minimum and not update $\beta_{\txt{New}}$ too aggressively by taking a smaller value of $\Delta \beta$.
The latter causes overshooting of the minimum and prevents convergence of
$\beta_{\txt{New}}$ and, thus, minimization of the residual. On the other hand, if
convergence is reached the subroutine \textproc{computeRAW} applies the
Robert–Asselin–Williams (RAW) time filter
\citep{kwizak1971,williams2009,williams2011} to prevent instabilities associated
with time step size. The filter parameters are $\nu = 0.12$ and $\alpha = 0.5$.
After exiting the while loop, $\beta_{\txt{Now}}$ and $\beta_{\txt{Old}}$ are
both updated for the next time step.


The \textproc{computeArrayC} subroutine shown in Algorithm~\ref{alg:c} computes
the collision operator $C$. To start with, the PDF $f$ is
computed by taking the reverse Fourier transform of $\beta_{\txt{Now}}$. The
PDF is then used to compute the number density  $n_{\txt{Eq}}$
and the average velocities $U$, $V$, and $W$ as outlined in
section~2. The next step is to compute the squared velocity
fluctuations $UU$, $VV$, and $WW$. This is done in a non-nested loop to allow
for the parallel computation of these variables. Afterward the
\textproc{reshape} function is used to cast these arrays in a $N \times N \times
N \times N$ form, where $N$ is the number of collocation points for each
dimension. Afterwards the squared velocity fluctuation fields are decomposed
using CANDECOMP/PARAFAC (CP) decomposition \citep{kolda2009} in the
\textproc{decompose} subroutine. The decomposition goes significantly faster if
an initial guess is chosen to be close to the final answer. Therefore the
decomposition result of the previous time step is passed on as the initial guess
when calling the \textproc{decompose} subroutine. Using the decomposed velocity
fluctuation fields, the temperature and collision frequency are computed in
\textproc{computeTemperature} and \textproc{computeFrequency}, respectively.

The same procedure is followed to compute the exponential blocks of the
collision operator $\txt{Exp}_{U}$, $\txt{Exp}_{V}$, and $\txt{Exp}_{W}$. Before
multiplying these different blocks they need to be projected on the 6D phase space
of the collision operator. This is done using the \textproc{kreshape}
subroutine. $\txt{Exp}_{U}$ is projected onto the ($\mathbf x,\xi_1$) space,
$\txt{Exp}_{V}$ onto the ($\mathbf x,\xi_2$) space, and $\txt{Exp}_{W}$ onto
the ($\mathbf x,\xi_3$) space. After constructing the equilibrium distribution by
multiplying $\txt{Exp}_{U}$, $\txt{Exp}_{V}$, and $\txt{Exp}_{W}$, the
difference between the equilibrium and the actual distribution is computed and
the result is multiplied with the collision frequency $\nu$ and divided by the
Knudsen number $\txt{Kn}$.

The \textproc{computeArrayN0} subroutine in Algorithm~\ref{alg:n0}
computes array $N_{0}$, which is part of the set of equations that solves the
Boltzmann transport equation. First the operators $A^{-}$ and $A^{+}$ are
applied to $\beta_{\txt{Old}}$ and $\beta_{\txt{New}}$, respectively, using
subroutine \textproc{computeG}. The reason $\beta_{\txt{New}}$ is repeated
$R_{\txt{Exp}}$ times is that eventually $\beta_{\txt{New}}$ is combined with
the Kruskal operator $\txt{Exp}$ with rank $R_{\txt{Exp}}$, which is a representation of trial function
$\phi_{s}$. After operators $A^{-}$ and  $A^{+}$ are applied,
$\beta_{\txt{Old}}$ and $\beta_{\txt{New}}$ are multiplied with each other. The
\textproc{mtimeskBlock} subroutine is used to make sure that both are multiplied
one rank at a time. The next step performs integration for each dimension. This is possible because decomposition has made the
different dimensions independent of each other. The next block of code
incorporates the trial function multiplied with $A^{+}$ into the subroutine. For
each dimension the Kruskal operator $\txt{Exp}_{A^{+}} \brc{d}$ is
two-dimensional and $\beta_{\txt{Old}}$ is reshaped to match this before
multiplying the two using \textproc{mtimeskBlock}. The next two lines integrate over
the first dimension so only the wave number is left as a variable. The last
block of code combines the results of the above shown calculations. The Kruskal
operator $\txt{Exp}_{A^{+}} \brc{d}$ is only used when solving for
$\beta_{\txt{New}} \brc{d}$. Therefore, for every dimension $\txt{Exp}_{A^{+}}
\brc{d}$ is multiplied by  $\beta_{\txt{New}} \brc{dd}$ when $d \ne dd$.
Finally, the results are summed up for every rank of $\beta_{\txt{New}}$, the
output is reshaped into one vector for every dimension, and $N$ is returned by the
subroutine.

\begin{algorithm}[htbp]
\caption{computeArrayN0 subroutine}
\label{alg:n0}
\begin{algorithmic}[1]
\Function{computeArrayN0}{$A^{+}, A^{-}, \txt{Exp}_{A^{+}},\beta_{\txt{Old}},\beta_{\txt{New}}$}
\State{$\repl{\beta_{\txt{New}}}{\beta_{\txt{Old}}} \gets \Call{computeG}{A^{-},\beta_{\txt{Old}},R_{\txt{Old}},1}$} \Comment{multiply with $A^{-}$ and iFFT}
\State{$\repl{\beta_{\txt{New}}}{\beta_{\txt{New}}} \gets \Call{computeG}{A^{+},\beta_{\txt{New}},R_{\txt{New}},R_{\txt{Exp}}}$} \Comment{multiply with $A^{+}$ and iFFT}
\State{$\beta_{\txt{New}} \gets \Call{mtimeskBlock}{\beta_{\txt{Old}},\beta_{\txt{New}},R_{\txt{Old}},R_{\txt{New}},4,4 \; R_{\txt{Exp}}}$} \Comment{Multiply per block}
\State
\For{$d \gets 1:6$}
\State{$\beta_{\txt{New}} \brc{d} \gets \brc{2 \pi}/N \; \Call{sum}{\beta_{\txt{New}} \brc{d},1}$} \Comment{Integrate all dimensions individually}
\EndFor
\State
\For{$d \gets 1:6$}
\State{$\beta_{\txt{Old},\txt{1-D}} \gets \Call{ktensor}{\beta_{\txt{Old}} \brc{d}}$}
\State{$\beta_{\txt{Old},\txt{1-D}} \gets
\Call{kreshape}{\beta_{\txt{Old},\txt{1-D}},[1 \; 0]}$} \Comment{Match dimension of $\txt{Exp}_{A^{+}} \brc{d}$}
\State
\State{$\txt{Exp}_{A^{+}} \brc{d} \gets \Call{mtimeskBlock}{\beta_{\txt{Old},\txt{1-D}},\txt{Exp}_{A^{+}} \brc{d},R_{\txt{Old}},1,4,4 \; R_{\txt{Exp}}}$}
\State
\State{$\txt{Exp}_{A^{+}} \brc{d} \brc{1} \gets \brc{2 \pi}/N \; \Call{sum}{\txt{Exp}_{A^{+}} \brc{d} \brc{1},1}$}
\State{$\repl{\txt{Exp}_{A^{+}} \brc{d} \brc{1}}{V \brc{d}} \gets
\txt{Exp}_{A^{+}} \brc{d} \brc{2} \cdot \txt{Exp}_{A^{+}} \brc{d} \brc{1}$} \Comment{Integrate over first dimension}
\EndFor
\State
\State{$N \gets \Call{ktensor}{V}$}
\State{$R_{\txt{N}} \gets 4 R_{\txt{Exp}} \; 4 R_{\txt{Old}}$}
\For{$d \gets 1:6$}
\State{$U \brc{d} \gets \Call{repmat}{N \brc{d},1,R_{\txt{New}}}$} \Comment{Repeat for each rank of beta}
\For{$dd \gets 1:6$}
\If{$d \sim = dd$} \Comment{Trial function is only used when $d = dd$}
\State{$U \brc{d} \gets U \brc{d} \cdot \beta_{\txt{New}} \brc{dd}$}
\EndIf
\EndFor
\For{$r \gets 1:R_{\txt{New}}$} \Comment{Summation for every rank of $\beta_{\txt{New}}$}
\State{$r_{0} \gets R_{N} (r-1) +1$}
\State{$r_{1} \gets R_{N} \; r    $}
\State{$U \brc{d}(:,r) \gets \Call{sum}{U \brc{d}(:,r_{0}:r_{1}),2}$}
\EndFor
\State{$U \brc{d} \gets U \brc{d}(:,1:R_{\txt{New}})$}
\State{$U \brc{d} \gets \Call{reshape}{U \brc{d},[],1}$} \Comment{Turn into one long array}
\EndFor
\State{$N \gets U$}
\State \Return{$N$}
\EndFunction
\end{algorithmic}
\end{algorithm}


Subroutine \textproc{computeArrayO0} in Algorithm~\ref{alg:o0} adds the
contribution of the collision operator $C$ to 
the Boltzmann transport equation. First, $\beta_{\txt{New}}$ is expanded to
match the rank of $\txt{Exp}_{A^{+}}$, operator $A^{+}$ is applied to
$\beta_{\txt{New}}$, and the inverse Fourier transform is taken using subroutine
\textproc{computeG}. Afterward, every rank of $\beta_{\txt{New}}$ is multiplied
with $C$ and each dimension is integrated over. The next block of code deals
with matching the dimensions of  $C$ with $\txt{Exp}_{A^{+}}$, multiplying the
two, and integrating over the real space variable so only the wave number
remains. In the last part of the subroutine the results are combined again. In Algorithm~\ref{alg:n0}, when $d$ matches the dimension that is solved for, $\txt{Exp}_{A^{+}} \brc{d}$ is used and otherwise $\beta_{\txt{New}}
\brc{d}$ is used in the multiplication. Lastly, the output is reshaped into one
vector for every dimension, and $O$ is returned by the subroutine.

\begin{algorithm}[H]
\caption{computeArrayO0 subroutine}
\label{alg:o0}
\begin{algorithmic}[1]
\Function{computeArrayO0}{$C,A^{+},\txt{Exp}_{A^{+}},\beta_{\txt{New}}$}
\State{$\beta_{\txt{New}} \gets \Call{computeG}{A^{+},\beta_{\txt{New}},R_{\txt{New}},R_{\txt{Exp}}}$} \Comment{multiply with $A^{+}$ and iFFT}
\State
\State{$\beta_{\txt{New}} \gets \Call{mtimeskBlock}{C,\beta_{\txt{New}},1,R_{\txt{New}},R_{\txt{C}},4 \; R_{\txt{Exp}}}$} \Comment{Multiply per rank}
\State
\For{$d \gets 1:6$}
\State{$\beta_{\txt{New}} \brc{d} \gets \brc{2 \pi}/N \; \Call{sum}{\beta_{\txt{New}} \brc{d},1}$} \Comment{Integrate over each dimension}
\EndFor
\State
\For{$d \gets 1:6$}
\State{$C_{\txt{1-D}} \gets \Call{ktensor}{C \brc{d}}$}
\State{$C_{\txt{1-D}} \gets \Call{kreshape}{C_{\txt{1-D}},[1 \; 0]}$} \Comment{Match dimension of $\txt{Exp}_{A^{+}} \brc{d}$}
\State
\State{$\txt{Exp}_{A^{+}} \brc{d} \gets \Call{mtimesk}{C_{\txt{1-D}}, \txt{Exp}_{A^{+}} \brc{d}}$}
\State{$\txt{Exp}_{A^{+}} \brc{d} \brc{1} \gets  \brc{2 \pi}/N \; \Call{sum}{\txt{Exp}_{A^{+}} \brc{d} \brc{1},1}$}
\State{$U \brc{d} \gets \txt{Exp}_{A^{+}} \brc{d} \brc{2} \cdot \txt{Exp}_{A^{+}} \brc{d} \brc{1}$} \Comment{Only the wave number remains}
\EndFor
\State
\State{$O \gets \Call{ktensor}{U}$}
\State{$R_{\txt{O}} \gets 4 R_{\txt{Exp}} \; R_{\txt{c}}$}
\State
\For{$d \gets 1:6$}
\State{$U \brc{d} \gets \Call{repmat}{O \brc{d},1,R_{\txt{New}}}$} \Comment{Repeat for each rank of beta}
\For{$dd \gets 1:6$}
\If{$d \sim = dd$} \Comment{Trial function is only used when $d = dd$}
\State{$U \brc{d} \gets U \brc{d} \cdot \beta_{\txt{New}} \brc{dd}$}
\EndIf
\EndFor
\State
\For{$r \gets 1:R_{\txt{New}}$} \Comment{Summation for every rank of $\beta_{\txt{New}}$}
\State{$r_{0} = R_{O} (r-1) +1$}
\State{$r_{1} = R_{O} \; r    $}
\State{$U \brc{d}(:,r) \gets \Call{sum}{U \brc{d}(:,r_{0}:r_{1}),2}$}
\EndFor
\State
\State{$U \brc{d} \gets U \brc{d}(:,1:R_{\txt{New}})$}
\State{$U \brc{d} \gets \Call{reshape}{U \brc{d},[],1}$} \Comment{Turn into one long array}
\EndFor
\State
\State{$O \gets U$}
\State
\State \Return{$O$}
\EndFunction
\end{algorithmic}
\end{algorithm}

Subroutine \textproc{computeArrayM0} in Algorithm~\ref{alg:m0} is the last
matrix to be computed in order to solve the Boltzmann equation. This
subroutine is parallelized and the dimension $d$ is passed to the
subroutine to indicate for which dimension array $M_{0}$ is computed. The code 
starts with subroutine \textproc{computeG} expanding $\beta_{\txt{New}}$ to
match the rank of $\txt{Exp}_{A^{+}}$, applying operator $A^{+}$ to
$\beta_{\txt{New}}$, and the taking the inverse Fourier transform. In subroutine
\textproc{computeInt}, $\beta_{\txt{New}}$ is multiplied with itself rank by rank
and then integration is performed over each dimension. Since $M_{0}$ is a
2D array with one wave number for each dimension,
$\txt{Exp}_{A^{+}}$ is expanded to three dimensions with two 
mappings, $\txt{Exp}_{A^{+},2}$ and $\txt{Exp}_{A^{+},3}$. These mappings are
multiplied with each other and then integration is performed over the first
dimension to reduce the operator from 3D to 2D. The
last part of the algorithm maps the different rank combinations, which result
from the multiplication of $\beta_{\txt{New}}$ with itself, onto different parts of
array $M_{0}$. Then $M_{0}$ is returned by the subroutine.

\begin{algorithm}[htbp]
\caption{computeArrayM0 subroutine}
\label{alg:m0}
\begin{algorithmic}[1]
\Function{computeArrayM0}{$A^{+},\txt{Exp}_{A^{+}},\beta_{\txt{New}},d$}
\State{$\beta_{\txt{New}} \gets \Call{computeG}{A^{+},\beta_{\txt{New}},R_{\txt{New}},R_{\txt{Exp}}}$} \Comment{multiply with $A^{+}$ and iFFT}
\State{$\beta_{\txt{New}} \gets \Call{computeInt}{\beta_{\txt{New}}}$} \Comment{Perform integration}
\State{$\txt{Exp}_{A^{+},2} \gets \Call{kreshape}{\txt{Exp}_{A^{+}},[1 \; 1 \; 0] }$} \Comment{Turn into 3D for different wave numbers}
\State{$\txt{Exp}_{A^{+},3} \gets \Call{kreshape}{\txt{Exp}_{A^{+}},[1 \; 0 \; 1] }$}
\State{$\txt{Exp}_{A^{+}} \gets \Call{mtimesk}{\txt{Exp}_{A^{+},2},\txt{Exp}_{A^{+},3}}$}
\State{$\txt{Exp}_{A^{+}} \brc{1} \gets (2 \pi/N) \; \Call{sum}{\txt{Exp}_{A^{+}} \brc{1},1}$} \Comment{Integrate over real space}
\For{$dd \gets 1:2$} \Comment{Map to two dimensions}
\State{$U \brc{dd} \gets \txt{Exp}_{A^{+}} \brc{dd+1} \cdot \sqrt{\txt{Exp}_{A^{+}} \brc{1}}$}
\EndFor
\State{$M \gets \Call{ktensor}{U}$}
\State{$M \gets \Call{krepmat}{M,R_{\txt{New}}^{2}}$} \Comment{Repeat array to match rank of $\beta_{\txt{New}}$}
\For{$dd \gets 1:6$} \Comment{Trial function is only used when $d = dd$}
\If{$d \sim = dd$}
\State{$M \brc{d} \gets M \brc{d} \cdot \beta_{\txt{New}} \brc{dd}$}
\EndIf
\EndFor
\State{$R_{M} \gets 4 \; R_{\txt{Exp}} \; 4 \; R_{\txt{Exp}}$} \Comment{Each rank combination is a subarray of $M_{0}$}
\For{$r \gets 1:R_{\txt{New}}^{2}$}
\State{$r_{0} \gets R_{M} (r-1)+1$}
\State{$r_{1} \gets R_{M} \; r   $}
\For{$dd \gets 1:2$}
\State{$U \brc{dd} = M \brc{dd} (:,r_{0}:r_{1})$}
\EndFor
\State{$x \gets \Call{mod}{r-1,R_{\txt{New}}}+1$} \Comment{Compute coordinates of rank combination}
\State{$y \gets \brc{r-x}/R_{\txt{New}}+1$}
\State{$x_{0} \gets N (x-1)+1$}
\State{$x_{1} \gets N \; x   $}
\State{$y_{0} \gets N (y-1)+1$}
\State{$y_{1} \gets N \; y   $}
\State{$M_{\txt{tmp}} (x_{0}:x_{1},y_{0}:y_{1}) \gets \Call{double}{}\brc{\Call{ktensor}{U}}$}
\EndFor
\State{$M \gets M_{\txt{tmp}}$}
\State \Return{$M$}
\EndFunction
\end{algorithmic}
\end{algorithm}

The next two subroutines \textproc{computeArrayN1} and \textproc{computeArrayM1}
in Algorithms~\ref{alg:n1} and~\ref{alg:m1}, respectively, are used to enforce
continuity to the solution of the Boltzmann transport equation. Subroutine
\textproc{computeArrayN1} starts with computing the density field,
$n_{\txt{Eq,Now}}$. \textproc{computeF} takes the inverse Fourier transform and
computes the PDF, \textproc{kintegrate} performs
integration over velocity space, and \textproc{kreshape} maps the 3D density
field onto four dimensions for multiplication later in the code. The advection
term $nU_{\txt{Now}}$ is computed next. \textproc{computeG} applies operator $A$
and takes the inverse Fourier transform, \textproc{kintegrate} performs
integration over velocity space, and \textproc{kreshape} maps the 3D field onto
four dimensions again. The next sections of the code involve multiplication of
$n_{\txt{Eq,New}}$ with $n_{\txt{Eq,Now}}$, $n_{\txt{Eq,New}}$ with
$nU_{\txt{Now}}$, $nU_{\txt{New}}$ with $n_{\txt{Eq,Now}}$, and $nU_{\txt{New}}$
with $nU_{\txt{Now}}$. For all these terms the derivative with respect to the
different dimensions of $\beta_{\txt{New}}$ is taken. For the first term
representing the derivative of $n_{\txt{Eq,New}} \cdot n_{\txt{Eq,Now}}$,
$\beta_{\txt{New}}$ is expanded, $f_{\txt{New}}$ is computed, and
$\txt{Exp}_{0}$ is repeated $R_{\txt{New}}$ times to match the rank of these
operators for multiplication. Now, for all dimensions that are not equal to the
dimension being resolved the trial function operator $\txt{Exp}$ is multiplied
with $\beta_{\txt{New}}$. In the next step the resulting seven dimensional field
is integrated over the velocity phase space and multiplied with the density
field $n_{\txt{Eq,Now}}$. After multiplication the resulting field is also
integrated over position space resulting in a 1Dvector that only
depends on the wave number. The last loop sums over the vector for every rank of
$\beta_{\txt{New}}$ and transforms the data into one long vector. The next
sections of the code that deal with the other multiplications work in the same
way and at the end of the subroutine $N_{1}$ is returned.

\begin{algorithm}[htbp]
\caption{computeArrayN1 subroutine}
\label{alg:n1}
\begin{algorithmic}[1]
\Function{computeArrayN1}{$\txt{Exp},A,\txt{Exp}_{A},\beta_{\txt{Now}},\beta_{\txt{New}}$}
\State{$n_{\txt{Eq,Now}} \gets \Call{computeF}{\beta_{\txt{Now}}, [1 \; 1 \; 1 \; 1 \; 1 \; 1]}$} \Comment{Compute $f_{\txt{Now}}$}
\State{$n_{\txt{Eq,Now}} \gets \Call{kintegrate}{n_{\txt{Eq,Now}}, [0 \; 0 \; 0 \; 1 \; 1 \; 1]}$} \Comment{Compute $n_{\txt{Eq,Now}}$}
\State{$n_{\txt{Eq,Now}} \gets \Call{kreshape}{n_{\txt{Eq,Now}},[1 \; 1 \; 1 \; 0]}$} \Comment{Add dimension}
\State{$nU_{\txt{Now}} \gets
\Call{computeG}{A,\beta_{\txt{Now}},R_{\txt{Now}},1}$} \Comment{Apply A to $\beta_{\txt{Now}}$ and iFFT}
\State{$nU_{\txt{Now}} \gets \Call{kintegrate}{nU_{\txt{Now}}, [0 \; 0 \; 0 \; 1 \; 1 \; 1]}$} \Comment{Integrate}
\State{$nU_{\txt{Now}} \gets \Call{kreshape}{nU_{\txt{Now}},[1 \; 1 \; 1 \; 0]}$} \Comment{Add dimension}
\State{$\beta_{\txt{New},0} \gets
\Call{krepelem}{\beta_{\txt{New}},R_{\txt{Exp}}}$} \Comment{Expand $\beta_{\txt{New}}$ to rank $R_{\txt{Exp}}$}
\State{$\beta_{\txt{New},0} \gets \Call{computeF}{\beta_{\txt{New},0},[1 \; 1 \; 1 \; 1 \; 1 \; 1]}$} \Comment{Compute $f_{\txt{New}}$}
\State{$\txt{Exp}_{0} \gets \Call{krepmat}{\txt{Exp},R_{\txt{New}}}$} \Comment{Repeat $R_{\txt{New}}$ times}
\For{$d \gets 1:6$}
\State{$V \brc{d} \gets \txt{Exp}_{0} \brc{1}$} \Comment{Assign real variable to dimension $d$}
\State{$V \brc{7} \gets \txt{Exp}_{0} \brc{2}$} \Comment{Assign wave number to additional dimension}
\For{$dd \gets 1:6$} \Comment{Trial function is only used when $d = dd$}
\If{$d \sim = dd$}
\State{$V \brc{dd} \gets \beta_{\txt{New},0} \brc{dd}$}
\EndIf
\EndFor
\State{$N \gets \Call{ktensor}{V}$}
\State{$N \gets \Call{kintegrate}{N,[0 \; 0 \; 0 \; 1 \; 1 \; 1 \; 0]}$} \Comment{Integrate over velocity}
\State{$N \gets \Call{mtimeskBlock}{n_{\txt{Eq,Now}},N,R_{\txt{Now}},R_{\txt{New}},1,R_{\txt{Exp}}}$} \Comment{Multiply}
\State{$N \gets \Call{kintegrate}{N,[1 \; 1 \; 1 \; 0]}$} \Comment{Integrate over positions}
\State{$R_{N} \gets R_{\txt{Exp}} \; R_{\txt{Now}}$}
\For{$r \gets 1:R_{\txt{New}}$} \Comment{Summation for every rank of $\beta_{\txt{New}}$}
\State{$r_{0} \gets R_{N} (r-1)+1$}
\State{$r_{1} \gets R_{N}  \; r  $}
\State{$U \brc{d} (:,r) \gets \Call{sum}{N \brc{1} (:,r_{0}:r_{1}),2}$}
\EndFor
\State{$U \brc{d} \gets U \brc{d} (:,1:R_{\txt{New}})$} \Comment{After summation array is stored in $(:,1:R_{\txt{New}})$}
\State{$U \brc{d} \gets \Call{reshape}{U \brc{d},[],1}$}
\EndFor   
\State{$N_{0} \gets U$}
\State{$\beta_{\txt{New},0} \gets \Call{krepelem}{\beta_{\txt{New}},R_{\txt{Exp}}}$} \Comment{Expand $\beta_{\txt{New}}$ and $\txt{Exp}$}
\algstore{myalg}
\end{algorithmic}
\end{algorithm}

\begin{algorithm}[H]
\begin{algorithmic}[1]
\algrestore{myalg}
\State{$\beta_{\txt{New},0} \gets \Call{computeF}{\beta_{\txt{New},0},[1 \; 1 \; 1 \; 1 \; 1 \; 1]}$}
\State{$\txt{Exp}_{0} \gets \Call{krepmat}{\txt{Exp},R_{\txt{New}}}$}
\For{$d \gets 1:6$}
\State{$V \brc{d} \gets \txt{Exp}_{0} \brc{1}$} \Comment{Assign real variable to dimension $d$}
\State{$V \brc{7} \gets \txt{Exp}_{0} \brc{2}$} \Comment{Assign wave number to additional dimension}
\For{$dd \gets 1:6$} \Comment{Trial function is only used when $d = dd$}
\If{$d \sim = dd$}
\State{$V \brc{dd} \gets \beta_{\txt{New},0} \brc{dd}$}
\EndIf
\EndFor
\State{$N \gets \Call{ktensor}{V}$}
\State{$N \gets \Call{kintegrate}{N,[0 \; 0 \; 0 \; 1 \; 1 \; 1 \; 0]}$} \Comment{Integrate over velocity}
\State{$N \gets \Call{mtimeskBlock}{nU_{\txt{Now}},N,R_{\txt{Now}},R_{\txt{New}},3,R_{\txt{Exp}}}$} \Comment{Multiply}
\State{$N \gets \Call{kintegrate}{N,[1 \; 1 \; 1 \; 0]}$} \Comment{Integrate over positions}
\State{$R_{N} \gets R_{\txt{Exp}} \; 3 \; R_{\txt{Now}}$}
\For{$r \gets 1:R_{\txt{New}}$} \Comment{Summation for every rank of $\beta_{\txt{New}}$}
\State{$r_{0} \gets R_{N} (r-1)+1$}
\State{$r_{1} \gets R_{N}  \; r  $}
\State{$U \brc{d} (:,r) \gets \Call{sum}{N \brc{1} (:,r_{0}:r_{1}),2}$}
\EndFor
\State{$U \brc{d} \gets U \brc{d} (:,1:R_{\txt{New}})$} \Comment{After summation array is stored in $(:,1:R_{\txt{New}})$}
\State{$U \brc{d} \gets \Call{reshape}{U \brc{d},[],1}$}
\EndFor   
\For{$d \gets 1:6$}
\State{$N_{0} \brc{d} \gets N_{0} \brc{d} - U \brc{d}$}
\EndFor   
\State{$\beta_{\txt{New},0} \gets \Call{computeG}{A,\beta_{\txt{New}},R_{\txt{New}},R_{\txt{Exp}}}$} \Comment{Expand $\beta_{\txt{New}}$ and $\txt{Exp}$}
\For{$d \gets 1:6$}
\State{$\txt{Exp}_{A,0} \gets \Call{krepmat}{\txt{Exp}_{A} \brc{d},R_{\txt{New}}}$}
\State{$V \brc{d} \gets \txt{Exp}_{A,0} \brc{1}$} \Comment{Assign real variable to dimension $d$}
\State{$V \brc{7} \gets \txt{Exp}_{A,0} \brc{2}$} \Comment{Assign wave number to additional dimension}
\For{$dd \gets 1:6$} \Comment{Trial function is only used when $d = dd$}
\If{$d \sim = dd$}
\State{$V \brc{dd} \gets \beta_{\txt{New},0} \brc{dd}$}
\EndIf
\EndFor
\State{$N \gets \Call{ktensor}{V}$}
\State{$N \gets \Call{kintegrate}{N,[0 \; 0 \; 0 \; 1 \; 1 \; 1 \; 0]}$} \Comment{Integrate over velocity}
\State{$N \gets \Call{mtimeskBlock}{n_{\txt{Eq,Now}},N,R_{\txt{Now}},R_{\txt{New}},1,3 R_{\txt{Exp}}}$} \Comment{Multiply}
\State{$N \gets \Call{kintegrate}{N,[1 \; 1 \; 1 \; 0]}$} \Comment{Integrate over positions}
\State{$R_{N} \gets 3 \; R_{\txt{Exp}} \; R_{\txt{Now}}$}
\For{$r \gets 1:R_{\txt{New}}$} \Comment{Summation for every rank of $\beta_{\txt{New}}$}
\State{$r_{0} \gets R_{N} (r-1)+1$}
\State{$r_{1} \gets R_{N}  \; r  $}
\State{$U \brc{d} (:,r) \gets \Call{sum}{N \brc{1} (:,r_{0}:r_{1}),2}$}
\EndFor
\State{$U \brc{d} \gets U \brc{d} (:,1:R_{\txt{New}})$} \Comment{After summation array is stored in $(:,1:R_{\txt{New}})$}
\State{$U \brc{d} \gets \Call{reshape}{U \brc{d},[],1}$}
\EndFor   
\For{$d \gets 1:6$}
\State{$N_{0} \brc{d} \gets N_{0} \brc{d} + U \brc{d}$}
\EndFor
\State{$\beta_{\txt{New},0} \gets \Call{computeG}{A,\beta_{\txt{New}},R_{\txt{New}},R_{\txt{Exp}}}$} \Comment{Expand $\beta_{\txt{New}}$ and $\txt{Exp}$}
\algstore{myalg}
\end{algorithmic}
\end{algorithm}

\begin{algorithm}[H]
\begin{algorithmic}[1]
\algrestore{myalg}
\For{$d \gets 1:6$}
\State{$\txt{Exp}_{A,0} \gets \Call{krepmat}{\txt{Exp}_{A} \brc{d},R_{\txt{New}}}$}
\State{$V \brc{d} \gets \txt{Exp}_{A,0} \brc{1}$} \Comment{Assign real variable to dimension $d$}
\State{$V \brc{7} \gets \txt{Exp}_{A,0} \brc{2}$} \Comment{Assign wave number to additional dimension}
\For{$dd \gets 1:6$} \Comment{Trial function is only used when $d = dd$}
\If{$d \sim = dd$}
\State{$V \brc{dd} \gets \beta_{\txt{New},0} \brc{dd}$}
\EndIf
\EndFor
\State{$N \gets \Call{ktensor}{V}$}
\State{$N \gets \Call{kintegrate}{N,[0 \; 0 \; 0 \; 1 \; 1 \; 1 \; 0]}$} \Comment{Integrate over velocity}
\State{$N \gets \Call{mtimeskBlock}{nU_{\txt{Now}},N,R_{\txt{Now}},R_{\txt{New}},3,3 R_{\txt{Exp}}}$} \Comment{Multiply}
\State{$N \gets \Call{kintegrate}{N,[1 \; 1 \; 1 \; 0]}$} \Comment{Integrate over positions}
\State{$R_{N} \gets 3 \; R_{\txt{Exp}} \; 3 \; R_{\txt{Now}}$}
\For{$r \gets 1:R_{\txt{New}}$} \Comment{Summation for every rank of $\beta_{\txt{New}}$}
\State{$r_{0} \gets R_{N} (r-1)+1$}
\State{$r_{1} \gets R_{N}  \; r  $}
\State{$U \brc{d} (:,r) \gets \Call{sum}{N \brc{1} (:,r_{0}:r_{1}),2}$}
\EndFor
\State{$U \brc{d} \gets U \brc{d} (:,1:R_{\txt{New}})$} \Comment{After summation array is stored in $(:,1:R_{\txt{New}})$}
\State{$U \brc{d} \gets \Call{reshape}{U \brc{d},[],1}$}
\EndFor   
\For{$d \gets 1:6$}
\State{$N_{0} \brc{d} \gets N_{0} \brc{d} - U \brc{d}$}
\EndFor
\Return{$N_{0}$}
\EndFunction
\end{algorithmic}
\end{algorithm}

The last subroutine \textproc{computeArrayM1}, shown in Algorithm~\ref{alg:m1},
computes array $M_{1}$, the second and last array needed to enforce continuity.
Like subroutine \textproc{computeArrayM0}, this subroutine is also parallelized
and the dimension $d$ gets passed on to the subroutine to indicate for which
dimension array $M_{1}$ is computed. Like array $N_{1}$, this array also
consists of four blocks but this time representing: $n_{\txt{Eq,New}} \cdot
n_{\txt{Eq,New}}$, $n_{\txt{Eq,New}} \cdot nU_{\txt{New}}$, $nU_{\txt{New}}
\cdot n_{\txt{Eq,New}}$, and $nU_{\txt{New}} \cdot nU_{\txt{New}}$. The term
representing $nU_{\txt{New}}$ is computed using \textproc{computeG} and the term
representing $n_{\txt{Eq,New}}$ is computed using \textproc{krepelem} and
\textproc{computeF}. Because $M_{1}$ is a
2D array with one wave number for each dimension,
$\txt{Exp}$ is expanded to three dimensions with two different
mappings $\txt{Exp}_{2}$ and $\txt{Exp}_{3}$. These two mappings are appended
with the different components of $\beta_{\txt{New}} \brc{dd}$ for dimensions $d
\ne dd$. The next step is integration over velocity phase space, rank by rank multiplication
of the two resulting fields, and integration over position phase space to create
a two-dimensional array. The next part of the algorithm maps the different rank
combinations, which result from the multiplication $n_{\txt{Eq,New}} \cdot
n_{\txt{Eq,New}}$, onto different parts of array $M_{1}$. For the other three
blocks described above the same procedure is followed resulting in the complete
computation of array $M_{1}$, which is returned at the end of the subroutine.

\begin{algorithm}[htbp]
\caption{computeArrayM1 subroutine}
\label{alg:m1}
\begin{algorithmic}[1]
\Function{computeArrayM1}{$\txt{Exp},\txt{Exp}_{A},A,\beta_{\txt{New}},d$}
\State{$\beta_{\txt{New},0} \gets \Call{computeG}{A,\beta_{\txt{New}},R_{\txt{New}},R_{\txt{Exp}}}$} \Comment{Expand $\beta_{\txt{New}}$ and $\txt{Exp}$}
\State{$\beta_{\txt{New}} \gets \Call{krepelem}{\beta_{\txt{New}},R_{\txt{Exp}}}$} \Comment{Expand $\beta_{\txt{New}}$}
\State{$\beta_{\txt{New}} \gets \Call{computeF}{\beta_{\txt{New}},[1 \; 1 \; 1 \; 1 \; 1 \; 1]}$} \Comment{Transform into real space}
\algstore{myalg}
\end{algorithmic}
\end{algorithm}

\begin{algorithm}[H]
\begin{algorithmic}[1]
\algrestore{myalg}
\State{$\txt{Exp}_{2} \gets \Call{kreshape}{\txt{Exp},[1 \; 1 \; 0]}$} \Comment{Turn into 3-D arrays for different wave numbers}
\State{$\txt{Exp}_{3} \gets \Call{kreshape}{\txt{Exp},[1 \; 0 \; 1]}$}             
\State{$\txt{Exp}_{2} \gets \Call{krepmat}{\txt{Exp}_{2},R_{\txt{New}}}$} \Comment{Repeat $R_{\txt{New}}$ times}
\State{$\txt{Exp}_{3} \gets \Call{krepmat}{\txt{Exp}_{3},R_{\txt{New}}}$} 
\State{$U_{2} \brc{d} \gets \txt{Exp}_{2} \brc{1}$} \Comment{Assign real variable to dimension $d$}
\State{$U_{2} \brc{7} \gets \txt{Exp}_{2} \brc{2}$} \Comment{Assign wave number to additional dimension}
\State{$U_{2} \brc{8} \gets \txt{Exp}_{2} \brc{3}$}
\State{$U_{3} \brc{d} \gets \txt{Exp}_{3} \brc{1}$} \Comment{Assign real variable to dimension $d$}
\State{$U_{3} \brc{7} \gets \txt{Exp}_{3} \brc{2}$}
\State{$U_{3} \brc{8} \gets \txt{Exp}_{3} \brc{3}$} \Comment{Assign wave number to additional dimension}
\For{$dd \gets 1:6$} \Comment{Trial function is only used when $d = dd$}
\If{$d \sim = dd$}
\State{$U_{2} \brc{dd} \gets \beta_{\txt{New}} \brc{dd}$}
\State{$U_{3} \brc{dd} \gets \beta_{\txt{New}} \brc{dd}$}
\EndIf
\EndFor
\State{$M_{2} \gets \Call{ktensor}{U_{2}}$}
\State{$M_{3} \gets \Call{ktensor}{U_{3}}$}
\State{$M_{2} \gets \Call{kintegrate}{M_{2},[0 \; 0 \; 0 \; 1 \; 1 \; 1 \; 0 \; 0]}$} \Comment{Integrate over velocity}
\State{$M_{3} \gets \Call{kintegrate}{M_{3},[0 \; 0 \; 0 \; 1 \; 1 \; 1 \; 0 \; 0]}$}
\State{$M_{0} \gets
\Call{mtimeskBlock}{M_{2},M_{3},R_{\txt{New}},R_{\txt{New}},R_{\txt{Exp}},R_{\txt{Exp}}}$} \Comment{Multiplication}
\State{$M_{0} \gets \Call{kintegrate}{M_{0},[1 \; 1 \; 1 \; 0 \; 0]}$} \Comment{Integrate over positions}
\State{$R_{M} \gets R_{\txt{Exp}} \; R_{\txt{Exp}}$} \Comment{Each rank combination is a subarray of $M_{1}$}
\For{$r \gets 1:R_{\txt{New}}^{2}$}
\State{$r_{0} \gets R_{M} (r-1)+1$}
\State{$r_{1} \gets R_{M} \; r   $}
\For{$dd \gets 1:2$}
\State{$U \brc{dd} \gets M_{0} \brc{dd}(:,r_{0}:r_{1})$}
\EndFor
\State{$x \gets \Call{mod}{r-1,R_{\txt{New}}}+1$} \Comment{Compute coordinates of rank combination}
\State{$y \gets (r-x)/R_{\txt{New}}+1$}
\State{$x_{0} \gets N (x-1)+1$}
\State{$x_{1} \gets N \; x   $}
\State{$y_{0} \gets N (y-1)+1$}
\State{$y_{1} \gets N \; y   $}
\State{$M_{\txt{tmp}} (x_{0}:x_{1},y_{0}:y_{1}) \gets \Call{double}{} \; (\Call{ktensor}{U})$}
\EndFor
\State{$M \gets M_{\txt{tmp}}$}
\State{$\txt{Exp}_{2} \gets \Call{kreshape}{\repl{\txt{Exp}_{A}}{\txt{Exp}    } ,[1 \; 1 \; 0]}$} \Comment{Turn into 3-D arrays}
\State{$\txt{Exp}_{3} \gets \Call{kreshape}{\repl{\txt{Exp}_{A}}{\txt{Exp}_{A}} ,[1 \; 0 \; 1]}$}
\State{$\txt{Exp}_{2} \gets \Call{krepmat}{\txt{Exp}_{2},R_{\txt{New}}}$} \Comment{Repeat $R_{\txt{New}}$ times}
\State{$\txt{Exp}_{3} \gets \Call{krepmat}{\txt{Exp}_{3},R_{\txt{New}}}$} 
\State{$U_{2} \brc{d} \gets \txt{Exp}_{2} \brc{1}$} \Comment{Assign real variable to dimension $d$}
\State{$U_{2} \brc{7} \gets \txt{Exp}_{2} \brc{2}$} \Comment{Assign wave number to additional dimension}
\State{$U_{2} \brc{8} \gets \txt{Exp}_{2} \brc{3}$}
\State{$U_{3} \brc{d} \gets \txt{Exp}_{3} \brc{1}$} \Comment{Assign real variable to dimension $d$}
\State{$U_{3} \brc{7} \gets \txt{Exp}_{3} \brc{2}$}
\State{$U_{3} \brc{8} \gets \txt{Exp}_{3} \brc{3}$} \Comment{Assign wave number to additional dimension}
\algstore{myalg}
\end{algorithmic}
\end{algorithm}

\begin{algorithm}[H]
\begin{algorithmic}[1]
\algrestore{myalg}
\For{$dd \gets 1:6$} \Comment{Trial function is only used when $d = dd$}
\If{$d \sim = dd$}
\State{$U_{2} \brc{dd} \gets \repl{\beta_{\txt{New},0}}{\beta_{\txt{New}}  } \brc{dd}$}
\State{$U_{3} \brc{dd} \gets \repl{\beta_{\txt{New},0}}{\beta_{\txt{New},0}} \brc{dd}$}
\EndIf
\EndFor
\State{$M_{2} \gets \Call{ktensor}{U_{2}}$}
\State{$M_{3} \gets \Call{ktensor}{U_{3}}$}
\State{$M_{2} \gets \Call{kintegrate}{M_{2},[0 \; 0 \; 0 \; 1 \; 1 \; 1 \; 0 \; 0]}$} \Comment{Integrate over velocity}
\State{$M_{3} \gets \Call{kintegrate}{M_{3},[0 \; 0 \; 0 \; 1 \; 1 \; 1 \; 0 \; 0]}$}
\State
\State{$M_{0} \gets \Call{mtimeskBlock}{M_{2},M_{3},R_{\txt{New}},R_{\txt{New}},R_{\txt{Exp}},3 R_{\txt{Exp}}}$} \Comment{Multiplication}
\State{$M_{0} \gets \Call{kintegrate}{M_{0},[1 \; 1 \; 1 \; 0 \; 0]}$} \Comment{Integrate over positions}
\State{$R_{M} \gets R_{\txt{Exp}} \; 3 \; R_{\txt{Exp}}$} \Comment{Each rank combination is a subarray of $M_{1}$}
\For{$r \gets 1:R_{\txt{New}}^{2}$}
\State{$r_{0} \gets R_{M} (r-1)+1$}
\State{$r_{1} \gets R_{M} \; r   $}
\For{$dd \gets 1:2$}
\State{$U \brc{dd} \gets M_{0} \brc{dd}(:,r_{0}:r_{1})$}
\EndFor
\State{$x \gets \Call{mod}{r-1,R_{\txt{New}}}+1$} \Comment{Compute coordinates of rank combination}
\State{$y \gets (r-x)/R_{\txt{New}}+1$}
\State{$x_{0} \gets N (x-1)+1$}
\State{$x_{1} \gets N \; x   $}
\State{$y_{0} \gets N (y-1)+1$}
\State{$y_{1} \gets N \; y   $}
\State{$M_{\txt{tmp}} (x_{0}:x_{1},y_{0}:y_{1}) \gets \Call{double}{} \; (\Call{ktensor}{U})$}
\EndFor
\State{$M \gets M + M_{\txt{tmp}}$}
\State{$\txt{Exp}_{2} \gets \Call{kreshape}{\repl{\txt{Exp}_{A}}{\txt{Exp}_{A}}, [1 \; 1 \; 0]}$} \Comment{Turn into 3-D arrays}
\State{$\txt{Exp}_{3} \gets \Call{kreshape}{\repl{\txt{Exp}_{A}}{\txt{Exp}    }, [1 \; 0 \; 1]}$}
\State{$\txt{Exp}_{2} \gets \Call{krepmat}{\txt{Exp}_{2},R_{\txt{New}}}$}
\State{$\txt{Exp}_{3} \gets \Call{krepmat}{\txt{Exp}_{3},R_{\txt{New}}}$} 
\State{$U_{2} \brc{d} \gets \txt{Exp}_{2} \brc{1}$} \Comment{Assign real variable to dimension $d$}
\State{$U_{2} \brc{7} \gets \txt{Exp}_{2} \brc{2}$} \Comment{Assign wave number to additional dimension}
\State{$U_{2} \brc{8} \gets \txt{Exp}_{2} \brc{3}$}
\State{$U_{3} \brc{d} \gets \txt{Exp}_{3} \brc{1}$} \Comment{Assign real variable to dimension $d$}
\State{$U_{3} \brc{7} \gets \txt{Exp}_{3} \brc{2}$}
\State{$U_{3} \brc{8} \gets \txt{Exp}_{3} \brc{3}$} \Comment{Assign wave number to additional dimension}
\For{$dd \gets 1:6$} \Comment{Trial function is only used when $d = dd$}
\If{$d \sim = dd$}
\State{$U_{2} \brc{dd} \gets \beta_{\txt{New},0} \brc{dd}$}
\State{$U_{3} \brc{dd} \gets \beta_{\txt{New}} \brc{dd}$}
\EndIf
\EndFor
\State{$M_{2} \gets \Call{ktensor}{U_{2}}$}
\State{$M_{3} \gets \Call{ktensor}{U_{3}}$}
\State{$M_{2} \gets \Call{kintegrate}{M_{2},[0 \; 0 \; 0 \; 1 \; 1 \; 1 \; 0 \; 0]}$} \Comment{Integrate over velocity}
\State{$M_{3} \gets \Call{kintegrate}{M_{3},[0 \; 0 \; 0 \; 1 \; 1 \; 1 \; 0 \; 0]}$}
\State{$M_{0} \gets \Call{mtimeskBlock}{M_{2},M_{3},R_{\txt{New}},R_{\txt{New}},3 R_{\txt{Exp}},R_{\txt{Exp}}}$} \Comment{Multiplication}
\algstore{myalg}
\end{algorithmic}
\end{algorithm}

\begin{algorithm}[H]
\begin{algorithmic}[1]
\algrestore{myalg}
\State{$M_{0} \gets \Call{kintegrate}{M_{0},[1 \; 1 \; 1 \; 0 \; 0]}$} \Comment{Integrate over positions}
\State{$R_{M} \gets 3 \; R_{\txt{Exp}} \; R_{\txt{Exp}}$} \Comment{Each rank combination is a subarray of $M_{1}$}
\For{$r \gets 1:R_{\txt{New}}^{2}$}
\State{$r_{0} \gets R_{M} (r-1)+1$}
\State{$r_{1} \gets R_{M} \; r   $}
\For{$dd \gets 1:2$}
\State{$U \brc{dd} \gets M_{0} \brc{dd}(:,r_{0}:r_{1})$}
\EndFor
\State{$x \gets \Call{mod}{r-1,R_{\txt{New}}}+1$} \Comment{Compute coordinates of rank combination}
\State{$y \gets (r-x)/R_{\txt{New}}+1$}
\State{$x_{0} \gets N (x-1)+1$}
\State{$x_{1} \gets N \; x   $}
\State{$y_{0} \gets N (y-1)+1$}
\State{$y_{1} \gets N \; y   $}
\State{$M_{\txt{tmp}} (x_{0}:x_{1},y_{0}:y_{1}) \gets \Call{double}{} \; (\Call{ktensor}{U})$}
\EndFor
\State{$M \gets M + M_{\txt{tmp}}$}
\State{$\txt{Exp}_{2} \gets \Call{kreshape}{\txt{Exp}_{A},[1 \; 1 \; 0]}$} \Comment{Turn into 3-D arrays}
\State{$\txt{Exp}_{3} \gets \Call{kreshape}{\txt{Exp}_{A},[1 \; 0 \; 1]}$}
\State{$\txt{Exp}_{2} \gets \Call{krepmat}{\txt{Exp}_{2},R_{\txt{New}}}$}
\State{$\txt{Exp}_{3} \gets \Call{krepmat}{\txt{Exp}_{3},R_{\txt{New}}}$} 
\State{$U_{2} \brc{d} \gets \txt{Exp}_{2} \brc{1}$} \Comment{Assign real variable to dimension $d$}
\State{$U_{2} \brc{7} \gets \txt{Exp}_{2} \brc{2}$} \Comment{Assign wave number to additional dimension}
\State{$U_{2} \brc{8} \gets \txt{Exp}_{2} \brc{3}$}
\State{$U_{3} \brc{d} \gets \txt{Exp}_{3} \brc{1}$} \Comment{Assign real variable to dimension $d$}
\State{$U_{3} \brc{7} \gets \txt{Exp}_{3} \brc{2}$}
\State{$U_{3} \brc{8} \gets \txt{Exp}_{3} \brc{3}$} \Comment{Assign wave number to additional dimension}
\For{$dd \gets 1:6$} \Comment{Trial function is only used when $d = dd$}
\If{$d \sim = dd$}
\State{$U_{2} \brc{dd} \gets \beta_{\txt{New},0} \brc{dd}$}
\State{$U_{3} \brc{dd} \gets \beta_{\txt{New},0} \brc{dd}$}
\EndIf
\EndFor
\State{$M_{2} \gets \Call{ktensor}{U_{2}}$}
\State{$M_{3} \gets \Call{ktensor}{U_{3}}$}
\State{$M_{2} \gets \Call{kintegrate}{M_{2},[0 \; 0 \; 0 \; 1 \; 1 \; 1 \; 0 \; 0]}$} \Comment{Integrate over velocity}
\State{$M_{3} \gets \Call{kintegrate}{M_{3},[0 \; 0 \; 0 \; 1 \; 1 \; 1 \; 0 \; 0]}$}
\State{$M_{0} \gets \Call{mtimeskBlock}{M_{2},M_{3},R_{\txt{New}},R_{\txt{New}},3 R_{\txt{Exp}},3 R_{\txt{Exp}}}$} \Comment{Multiplication}
\State{$M_{0} \gets \Call{kintegrate}{M_{0},[1 \; 1 \; 1 \; 0 \; 0]}$} \Comment{Integrate over positions}
\State
\State{$R_{M} \gets 3 \; R_{\txt{Exp}} \; 3 \; R_{\txt{Exp}}$} \Comment{Each rank combination is a subarray of $M_{1}$}
\For{$r \gets 1:R_{\txt{New}}^{2}$}
\State{$r_{0} \gets R_{M} (r-1)+1$}
\State{$r_{1} \gets R_{M} \; r   $}
\For{$dd \gets 1:2$}
\State{$U \brc{dd} \gets M_{0} \brc{dd}(:,r_{0}:r_{1})$}
\EndFor
\State{$x \gets \Call{mod}{r-1,R_{\txt{New}}}+1$} \Comment{Compute coordinates of rank combination}
\State{$y \gets (r-x)/R_{\txt{New}}+1$}

\algstore{myalg}
\end{algorithmic}
\end{algorithm}

\begin{algorithm}[H]
\begin{algorithmic}[1]
\algrestore{myalg}
\State{$x_{0} \gets N (x-1)+1$}
\State{$x_{1} \gets N \; x   $}
\State{$y_{0} \gets N (y-1)+1$}
\State{$y_{1} \gets N \; y   $}
\State{$M_{\txt{tmp}} (x_{0}:x_{1},y_{0}:y_{1}) \gets \Call{double}{} \; (\Call{ktensor}{U})$}
\EndFor
\State{$M \gets M + M_{\txt{tmp}}$}
\Return{$M$}
\EndFunction
\end{algorithmic}
\end{algorithm}


\end{document}